\documentclass[aps,pra,twocolumn,superscriptaddress,longbibliography,nofootinbib]{revtex4-2}
\usepackage{amsfonts,amsmath,amssymb,epsfig}
\usepackage{epsfig}
\usepackage{graphicx}
\usepackage[dvipsnames]{xcolor}
\usepackage{color}
\usepackage{soul}
\usepackage{enumerate}
\usepackage{epstopdf}
\usepackage{array}
\usepackage{tabularx}
\usepackage[utf8]{inputenc}
\usepackage[T1]{fontenc}
\usepackage[colorlinks,linkcolor=blue, citecolor=blue,
urlcolor=blue]{hyperref}

\begin{document}
	\title{Equilibrium, Relaxation and Fluctuations in homogeneous Bose-Einstein Condensates: Linearized Classical Field Analysis}
	\author{Nils A. Krause}
	\email{krani857@student.otago.ac.nz}
	\affiliation{Department of Physics, University of Otago, Dunedin, New Zealand}
	\affiliation{Dodd-Walls Centre for Photonic and Quantum Technologies}
	\author{Ashton S. Bradley}
	\affiliation{Department of Physics, University of Otago, Dunedin, New Zealand}
	\affiliation{Dodd-Walls Centre for Photonic and Quantum Technologies}

	\date{\today}
	\begin{abstract}
		{Open quantum systems theory is central to describing the dynamics and equilibration of dilute-gas Bose-Einstein condensates (BECs).} We present an analysis of the linearized stochastic projected Gross-Pitaevskii equation (SPGPE) describing finite-temperature BECs. Our treatment provides an optimal choice for the cut-off that divides the Bose gas into the low-energy coherent region forming a classical wave, and the high-energy thermal cloud treated as a reservoir. Moreover, it highlights the relevance of energy damping, the number-conserving scattering between thermal and coherent atoms. We analyze the equilibrium properties and near-equilibrium relaxation of a homogeneous BEC in one, two and three dimensions at high phase-space density, and calculate the autocorrelation function and power spectrum of the density and phase fluctuations. Simulations of the full non-linear SPGPE are in close agreement, and extend our arguments beyond the linear regime. Our work suggests the need for a re-examination of decay processes in BECs studied under the neglect of energy damping. 
	\end{abstract}
	\flushbottom
	\maketitle
	\section{Introduction}
	{The theory of open quantum systems plays a foundational role in modern quantum physics, describing how quantum systems evolve when coupled to a thermal reservoir. This framework effectively captures Bose-Einstein condensation (BEC) from a quenched thermal cloud---including dynamics across the phase transition and the emergence of spontaneous vortices}~\cite{anderson_observation_1995, anglin_vortices_1999, weiler_spontaneous_2008}{. Energy exchange between the system and reservoir can induce sympathetic cooling}~\cite{myatt_production_1997}{, or drive internal convection in the superfluid}~\cite{gilz_quantum_2011, gilz_quantum_2015}{, underscoring the need to better understand reservoir interactions in dilute Bose gases. The advent of optical box traps}~\cite{gaunt_bose-einstein_2013, chomaz_emergence_2015, navon_quantum_2021}{ has enabled experimental exploration of homogeneous BECs, motivating a more rigorous treatment of BEC reservoir theory in such uniform systems.}

	In order to describe interacting Bose gases at finite temperature, various theories were developed that treat particles in highly occupied states collectively~\cite{zaremba_dynamics_1999,stoof_dynamics_2001,proukakis_finite-temperature_2008}. At high temperatures the stochastic projected Gross-Pitaevskii equation (SPGPE) emerged as a promising first-principles description of thermal dynamics~\cite{gardiner_stochastic_2002,gardiner_stochastic_2003,bradley_bose-einstein_2008,rooney_stochastic_2012,blakie_dynamics_2008}. In this formalism, particles occupying states above an energy cut-off are treated as a thermal reservoir, while particles below are treated as a classical field (c-field). Interactions between the reservoir and the c-field induce two damping mechanisms: an exchange of particles with the cloud called number damping, and scattering events leading to the exchange of energy (while conserving particle number) called energy damping.  
	
In this work, we employ the simplicity of the plane wave basis in homogeneous BECs to linearize the SPGPE around a coherent and homogeneous particle distribution. We perform a thorough analysis of the linearized SPGPE in one, two and three dimensions. In equilibrium at high phase-space density we find excellent agreement of the SPGPE theory with Bogoliubov theory, provided the thermal de Broglie wavelength of the particles is shorter than the superfluid healing length. We identify an optimal choice for the energy cut-off, thereby fixing the only free parameter of the theory. 

We demonstrate that energy damping is by far the dominant equilibration mechanism for most of the coherent region modes, and notably also for the near equilibrium condensate growth. Our work thus suggests that its inclusion dramatically alters the predicted dynamics and highlights the importance of energy-damping in open system dynamics. Moreover, our linearized approach enables calculating the autocorrelation of the density and phase fluctuations, as well as the full power spectrum analytically. The density autocorrelation provides a finite-temperature correction to the chemical potential. Full SPGPE simulations in two dimensions~\cite{bradley_low-dimensional_2015} confirm our analytical findings, and give insight about the breakdown of the linearization and physics beyond its validity.
	
	This work is structured as follows: in \ref{Background} we {discuss the literature concerning details of the SPGPE we address in this work, }state the full SPGPE and bring it into the dimensionless form used throughout. In \ref{Linearization} we perform the linearization and derive equations of motion for the density and phase of the coherent region. \ref{population} derives the single particle occupation according to the linearized SPGPE and compares it to the Bose-Einstein statistics assumed in the incoherent region as well as the Bogoliubov predictions. This analysis reveals an optimized cut-off choice as proposed in equation (\ref{cut-offchoice}). \ref{Decay} studies the decay of out-of-equilibrium population numbers, unveiling that thermalization of the coherent region in BECs is driven by energy damping. In \ref{timedependence} the autocorrelation and power spectrum are calculated. \ref{Chemical} presents a first order correction to the chemical potential. \ref{Numeric} gives details about the simulations performed to validate our analytical results. Finally, in \ref{Discussion} we present conclusions.
		
	\section{Background}
	\label{Background}
	C-field approximations rely on an additional free parameter, the energy cut-off $\epsilon_\text{cut}$ above which modes are treated as part of the incoherent thermal cloud. {In its choice it needs to be ascertained that the truncated Wigner approximation (TWA) can be employed. Validity of the TWA typically requires that} all states in the coherent region are \emph{high enough} occupied\cite{steel_dynamical_1998,norrie_quantum_2006}. While usually understood to mean the occupation should be larger than one, a quantitative estimate of the error made in such an approximation stays elusive. {In practice, there is no agreed upon way on fixing the cut-off. Instead, several approaches are employed, some of which are:}
	\begin{enumerate}[i)]
		\item {It is required that $N_\text{cut}\sim1-3$}\cite{bradley_low-dimensional_2015}{ and in particular $N_\text{cut}=1$}~\cite{comaron_quench_2019,roy_finite-temperature_2021,keepfer_phase_2022,roy_finite-temperature_2023,krause_thermal_2024}{. Here, $N_\text{cut}$ usually corresponds to its estimate according to the Bose-Einstein distribution. In a more sophisticated method attempting to better appreciate the influence of interactions, a cut-off is choosen according to the occupations calculated by the Hartree-Fock density of states. It is then demanded that there are as many single particle states in the c-region as there are in the Hartree-Fock theory below the choosen Hartree-Fock cut-off }\cite{rooney_decay_2010}.
		\item {$\epsilon_\text{cut}$ is required to be a few times the chemical potential $\epsilon_\text{cut}\sim2\mu-3\mu$}\cite{rooney_stochastic_2012,liu_dynamical_2018,liu_kibble-zurek_2020,mehdi_mutual_2023}{, so that the eigenstates of the Hamitlonian at and above the cut-off are basically the single particle states.}
		\item 	{A spatial grid is choosen and it is verified that the highest occupied state fulfils $N_\text{cut}>1$}\cite{groszek_crossover_2021,groszek_decaying_2020}.
	\end{enumerate}
	Typically, a check {of the validity} is only provided by verifying that a small change in the cut-off leads to only small changes in the predicted dynamics\cite{bradley_bose-einstein_2008}. {However, large variations usually violate SPGPE validity conditions.}
	
	Previous work on deriving an ideal cut-off aimed at describing the whole gas {within the c-field}~\cite{pietraszewicz_classical_2015,pietraszewicz_complex_2018,pietraszewicz_classical_2018,zawitkowski_classical-field_2004}. However, in the SPGPE a considerable fraction of the particles and kinetic energy are stored in the thermal cloud{, rendering these results ill suited}.   

	Due to its numerical complexity, the SPGPE is commonly reduced to an easier handleable simple-growth form {(called $\gamma$SPGPE in the following)}, neglecting the energy damping process{ and using spatially independent damping as derived} in \cite{bradley_bose-einstein_2008}. Application of this reduced equation succeeds in determining equilibrium states~\cite{bradley_bose-einstein_2008,keepfer_phase_2022,underwood_stochastic_2025} and gives qualitative insight into condensate dynamics. However, it fails in yielding quantitative dynamical results and the first-principles damping strength $\gamma$ is often enlarged or used as a fitting parameter\cite{weiler_spontaneous_2008,kobayashi_thermal_2016,liu_dynamical_2018,liu_kibble-zurek_2020,liu_vortex_2024,thudiyangal_universal_2024,comaron_quench_2019,groszek_crossover_2021}. Recent work highlighted the importance of the mechanism of energy damping in the decay of non-linear excitations\cite{mehdi_mutual_2023,krause_thermal_2024} motivating deeper study of its role in equilibration.

	{In the rest of this section }we briefly review the SPGPE reservoir theory and bring the equation into the dimensionless form used throughout this paper.
	
	\subsection{Stochastic projected Gross-Pitaevskii theory}
	The SPGPE theory\cite{gardiner_stochastic_2003,blakie_dynamics_2008} divides the atomic cloud into a dynamically evolving coherent region at low energies while high energetic atoms are treated as a reservoir. The latter high-energy region is assumed to stay thermally distributed at all times. In a (three-dimensional) box trap this leads to a thermal population of the state with momentum $\hbar\textbf{k}$
	\begin{align}
		\label{thoccupation}
			N_\textbf{k}^\text{th}&=\frac{1}{\exp(\beta[\epsilon(\textbf{k})-\mu_\text{3D}])-1},\\
		 \epsilon(\textbf{k})&=\frac{\hbar^2k^2}{2m}+2n_\text{tot}g.
	\end{align}
	Here, $m$ is the mass of the particles, $g=4\pi\hbar^2a_\text{s}/m$ the interaction strength, $a_\text{s}$ the s-wave scattering length, $\mu_{3\text{D}}$ the three dimensional chemical potential and $\beta=1/(k_\text{B}T)$ the inverse temperature. $n_\text{tot}=N_\text{tot}/L^3$ is the total density of the gas, where $N_\text{tot}$ is the total atom number and $L$ is the size of the box.
	
	The coherent region is described by an order parameter $\psi$ that evolves according to the Stratonovich equation of motion \cite{gardiner_stochastic_2003,rooney_stochastic_2012}
	\begin{subequations}
		\label{SPGPE}
		\begin{align}
			\label{fullSPGPE}
			\begin{split}
			i\hbar(\textbf{S})d\psi(\textbf{r})=&\mathcal{P}\left\{(\mathcal{L}_\text{GP}-\mu_\text{eff})\psi(\textbf{r})dt\right\}\\
			&+i\hbar d\psi(\textbf{r})|_\gamma+i\hbar (\textbf{S})d\psi(\textbf{r})|_\varepsilon.
			\end{split}
		\end{align}
		The first line contains the projected GPE with the Gross-Pitaevskii operator
		\begin{align}
			\mathcal{L}_\text{GP}\psi(\textbf{r})\equiv\left(-\frac{\hbar^2}{2m}\Delta+g|\psi(\textbf{r})|^2\right)\psi(\textbf{r}),
		\end{align}
		and the effective chemical potential $\mu_\text{eff}=\mu-2n_Ig${. The effective chemical potential contains the $n$-dimensional chemical potential $\mu$ (corresponding to the three-dimensional chemical potential $\mu_\text{3D}$ shifted by the ground state energy of the potential along the reduced dimensions }\cite{bradley_low-dimensional_2015}{) as well as the forward scattering Hamiltonian $2n_Ig$, where $n_I$ is the density of particles in the incoherent region. The first line} also features a projection on the low-energy region $\mathcal{P}$ that acts on a function $\phi$ as
		\begin{align}
			\mathcal{P}\{\phi(\textbf{r})\}=\int_{k<k_\text{cut}}\frac{d^n\textbf{k}}{(2\pi)^n}e^{i\textbf{k}\cdot\textbf{r}}\int d^n\textbf{r}'e^{-i\textbf{k}\cdot\textbf{r}'}\phi(\textbf{r}').
		\end{align}
		Here, $k_\text{cut}$ is the highest momentum in the coherent region. It relates to the cut-off energy $\epsilon_\text{cut}$ that divides the coherent and incoherent regions according to {(in three dimensions; in the two- and one-dimensionally reduced cases the cut-off energy is shifted by the ground state energy}\cite{bradley_low-dimensional_2015})
		\begin{align}
			\epsilon_\text{cut}=2n_\text{tot}g+\frac{\hbar^2k_\text{cut}^2}{2m}.
		\end{align}
		
		The second line in (\ref{fullSPGPE}) contains the damping and noise terms deriving from the interaction with the reservoir. Number damping and noise
		\begin{align}
			i\hbar d\psi(\textbf{r})|_\gamma&=\mathcal{P}\left\{-i\gamma(\mathcal{L}_\text{GP}-\mu_\text{eff})\psi(\textbf{r})dt+\sqrt{\frac{2\gamma k_\text{B}T}{\hbar}}dW\right\}\label{ndampdef}
			\end{align}
			and energy damping and noise
			\begin{align}
				\begin{split}
			i\hbar(\textbf{S})d\psi(\textbf{r})|_\varepsilon&=\mathcal{P}\bigg\{-\hbar\int d^n\textbf{r}'\varepsilon(\textbf{r}-\textbf{r}')\nabla'\cdot \textbf{j}(\textbf{r}')\psi(\textbf{r})dt\\
			&+\sqrt{\frac{2k_\text{B}T}{\hbar}}\psi(\textbf{r})\ dU(\textbf{r})\bigg\}.
			\end{split}
			\label{edampdef}
		\end{align}
		The Gaussian noise terms feature the correlations
		\begin{align}
			\langle dW^*(\textbf{r},t)dW(\textbf{r}',t)\rangle&=\delta(\textbf{r}-\textbf{r}')dt
		\end{align}
		and
		\begin{align}
			\langle dU(\textbf{r},t)dU(\textbf{r}',t)\rangle&=\varepsilon(\textbf{r}-\textbf{r}')dt.
		\end{align}
	\end{subequations}
	$\textbf{j}$ denotes the current density
	\begin{align}
		\textbf{j}=\frac{\hbar}{m}\text{Im}\left\{\psi^*\nabla\psi\right\}.
	\end{align}
	
	The number damping process (\ref{ndampdef}), can be characterized by a dimensionless damping strength~\cite{krause_thermal_2024}
	\begin{align}\label{gamdef}
		\gamma&=\frac{8a_\text{s}^2}{\lambda_\text{th}^2}e^{\beta\mu_{3\text{D}}}\int_0^1dy\ln\left(\frac{1-zy}{1-z}\right)\frac{1}{(1-y)(1-zy)},
	\end{align}
	where $z=e^{\beta(\mu_{3D}-2\epsilon_\text{cut})}<1$. We also introduced the thermal de Broglie wavelength
	\begin{align}
		\lambda_\text{th}=\sqrt{\frac{2\pi\hbar^2}{mk_\text{B}T}}.
	\end{align}
	Number damping captures those scattering events between particles in the reservoir in which one of the atoms loses enough energy to fall down into the coherent region, thereby driving growth and relaxation of the latter.
	
	The energy damping mechanism (\ref{edampdef}), is characterized by an integral kernel $\varepsilon(\mathbf{r})$. It describes the scattering between atoms in the coherent and incoherent region leading to a transfer of energy and thus damping excitations. Contrary to number damping, energy damping is number conserving. 
	The integral kernel in (\ref{edampdef})  depends on the spatial dimension. In the unreduced three dimensional case it is given by~\cite{rooney_stochastic_2012}
	\begin{align}
		\label{kernel}
		\varepsilon(\textbf{r})=16\pi a_\text{s}^2N_\text{cut}\int \frac{d^3\textbf{k}}{(2\pi)^3}\frac{e^{i\textbf{k}\cdot\textbf{r}}}{|\textbf{k}|}.
	\end{align}
	Here, $N_\text{cut}=(\exp[\beta (\epsilon_\text{cut}-\mu_{3D})]-1)^{-1}$ is the thermal population of modes at the cut-off energy. The expression for the kernel $\varepsilon$ in a reduced one- or two-dimensional system was found in \cite{bradley_low-dimensional_2015}. In the following we denote the kernel in $n$ dimensions as $\varepsilon_n$.
	
	\subsection{Dimensionless SPGPE}
	\label{dimensionless}
	In this section we re-write the SPGPE into a dimensionless form that will be convenient for performing the linearization.
	
	The SPGPE assumes the incoherent region to be thermalized at all times. In a homogeneous BEC it hence has the same density $n_I$ everywhere. We define the coherent region density
	\begin{align}
		n_C=n_\text{tot}-n_I.
	\end{align}
	Note that in out of equilibrium states the so defined coherent density can formally deviate from the average value of $|\psi|^2$ in the SPGPE theory.
	
	We perform the following scalings (with the healing length $\xi=\hbar/\sqrt{mn_Cg}$):
	\begin{align}
		\begin{split}
		&\textbf{r},\ \textbf{k}\rightarrow\xi\textbf{r},\ \textbf{k}/\xi,\ \psi\rightarrow\sqrt{n_C}\psi,\ t\rightarrow\hbar t/(n_Cg),\\ &dW\rightarrow\sqrt{\hbar/(n_Cg\xi^n)}dW,\ dU\rightarrow\sqrt{\frac{\xi^2\hbar}{n_Cg}}dU\\ &\varepsilon_n\rightarrow\xi^2\varepsilon_n,\ L\rightarrow L\xi.
		\end{split}
	\end{align}
	Further assuming a close to equilibrium state at high phase space density, we can set $\mu_\text{eff}\simeq n_Cg$, $n_Ig\simeq0$. We find the dimensionless SPGPE
		\begin{align}
			\label{dimensionlessSPGPE}
		\begin{split}
			i(\textbf{S})d\psi&=\mathcal{P}\bigg\{[1-i\gamma]\left[-\frac{\Delta}{2}+(|\psi|^2-1)\right]\psi dt\\
			&+\sqrt{\frac{4\pi}{n_C\lambda_\text{th}^2\xi^{n-2}}\gamma}dW-n_C\xi^n\int d^n\textbf{r}'\\
			&\times\left[\varepsilon_n(\textbf{r}-\textbf{r}')\nabla'\cdot\Im\{\psi^*\nabla'\psi\}\right]\psi(\textbf{r})dt\\
			&-\psi\sqrt{\text{{$\frac{4\pi\xi^2}{\lambda_\text{th}^2}$}}}dU\bigg\}.
		\end{split}
	\end{align}
	 The noise terms fulfill
	\begin{align}
		\begin{split}
			\langle dW^*(\textbf{r},t)dW(\textbf{r}',t)\rangle&=\delta(\textbf{r}-\textbf{r}')dt,\\
			\langle dU(\textbf{r},t)dU(\textbf{r}',t)\rangle&=\varepsilon_n(\textbf{r}-\textbf{r}')dt.
		\end{split}
	\end{align}
	
	The factor
	\begin{align}
		n_C\lambda_\text{th}^2\xi^{n-2}=n_C\lambda_\text{th}^n\left(\frac{\xi}{\lambda_\text{th}}\right)^{n-2}
	\end{align}
	determines the strength of the noise compared to the damping terms as well as the strength of energy damping compared to number damping. It is given by the $n$ dimensional phase space density scaled with the ratio of healing length and thermal de Broglie wavelength. Therefore, we will refer to it as the scaled phase space density in the following. As we will later see, use of the SPGPE requires $\xi>\lambda_\text{th}$. Thus, in three dimensions the noise increases with increasing temperature slower than the phase space density, while in one dimension its increase is enhanced, in agreement to the existence of a true condensate and a mere quasi-condensate, respectively.
	
	\subsection{Bogoliubov theory}
	We will benchmark the predictions of the linearized SPGPE against the known results of the Bogoliubov theory\cite{bogolubov_theory_1946,castin_bose-einstein_2001-1,mora_extension_2003}. Bogoliubov theory assumes that a large part of the particles ($N_0$ of the $N_\text{tot}$ particles) have entered the state with vanishing momentum, so that the annihilation and creation operators for this state fullfill
	\begin{align}
		\hat{a}_0\simeq\hat{a}_0^\dagger\simeq\sqrt{N_0}\simeq\sqrt{N_\text{tot}}.
	\end{align}
	The many particle Hamiltonian becomes
	\begin{align}
		\begin{split}
		\hat{H}&\simeq\frac{gn_\text{tot}}{2}N_\text{tot}+\sum_\textbf{k}\frac{\hbar^2k^2}{2m\xi^2}\hat{a}^\dagger_\textbf{k}\hat{a}_\textbf{k}\\
		&+\frac{n_\text{tot}g}{2}\sum_{k>0}\left[2\hat{a}_\textbf{k}^\dagger\hat{a}_\textbf{k}+\hat{a}_\textbf{k}^\dagger\hat{a}^\dagger_{-\textbf{k}}+\hat{a}_\textbf{k}\hat{a}_{-\textbf{k}}\right],
		\end{split}
	\end{align}
	where $\hat{a}_\textbf{k}$ and $\hat{a}_\textbf{k}^\dagger$ are the annihilation and creation operator for the single particle state with momentum $\textbf{k}$, fulfilling the commutation relation
    \begin{align}
        [\hat{a}_\textbf{k},\hat{a}^\dagger_{\textbf{k}'}]=\delta_{\textbf{k},\textbf{k}'}.
    \end{align}
	
	A Bogoliubov transformation is performed to find
	\begin{align}
		\hat{H}\simeq\frac{gn_\text{tot}}{2}N_\text{tot}+\sum_\textbf{k}\hbar\omega_k\hat{b}_\textbf{k}^\dagger\hat{b}_\textbf{k},
	\end{align}
	where, the quasi-particle creation and annihilation operators are given by
	\begin{align}
		\hat{b}_\textbf{k}=u_k\hat{a}_\textbf{k}-v_k\hat{a}^\dagger_{-\textbf{k}},\ u_k,v_k=\pm\sqrt{\frac{E_k}{2\omega_k}\pm\frac{1}{2}}.
	\end{align}
	Here, $E_k=1+k^2/2$ denotes the free particle energy and $\omega_k=\sqrt{k^2(1+k^2/4)}$ the Bogoliubov dispersion.
	
	The quasi-particles are Bose-Einstein distributed
	\begin{align}
		\langle\hat{b}^\dagger_\textbf{k}\hat{b}_\textbf{k}\rangle=\frac{1}{e^{\beta\hbar\omega_k}-1}
	\end{align}
	so that we have the occupation of the single-particle state with momentum $\textbf{k}$
	\begin{align}
		\label{Bogoliubovocc}
		\begin{split}
		N_\textbf{k}^B&=\langle\hat{a}_\textbf{k}^\dagger\hat{a}_\textbf{k}\rangle\\
		&=\frac{E_k/\omega_k}{\exp\left(\frac{\lambda_\text{th}^2}{2\pi\xi^2}\omega_k\right)-1}+\frac{1}{2}\left[\frac{E_k}{\omega_k}-1\right].
		\end{split}
	\end{align}
	In particular, we have for small momenta
	\begin{align}
		\label{Bogoliubovoccsmallk}
		N_\textbf{k}^B\sim2\pi\frac{\xi^2}{\lambda_\text{th}^2}\frac{1}{k^2}=\frac{mk_\text{B}T}{p^2}.
	\end{align}
	
	Having obtained the dimensionless form of the SPGPE we can now move on to calculate the equilibrium properties in the linearized regime.
	
	\section{Analytical results}
	In this section we calculate the equilibrium and linear equilibration properties of a Bose-Einstein condensate evolving according to the SPGPE. We will begin by linearizing the equation of motion (\ref{dimensionlessSPGPE}) under the use of a Madelung transform.
	
	\subsection{Linearization}
	\label{Linearization}
	If $\psi$ is only weakly disturbed from the homogeneous case, we can write (we work in the dimensionless form provided in \ref{dimensionless})
	\begin{align}
		\psi=\sqrt{1+\delta n}e^{i\theta}\simeq\left(1+\frac{\delta n}{2}\right)e^{i\theta}, \ n=1+\delta n.
	\end{align}
	Where density fluctuations are small $\delta n\ll1$. We assume coherence to hold over more than a healing length and hence expect a small phase gradient $\nabla\theta\ll1$. In lowest order the SPGPE then becomes (where we replace $dW\rightarrow e^{i\theta}dW$)
	\begin{align}
		\begin{split}
			&i(\textbf{S})d\left(\frac{\delta n}{2}+i\theta\right)\\
			&=\mathcal{P}\bigg\{[1-i\gamma]\left[-\frac{\Delta}{2}\left(\frac{\delta n}{2}+i\theta\right)+\delta n\right]dt\\
			&+\sqrt{\frac{4\pi\gamma}{n_C\lambda_\text{th}^2\xi^{n-2}}}dW\\
			&-\int d^n\textbf{r}'n_C\xi^{n}\varepsilon_n(\textbf{r}-\textbf{r}')\Delta'\theta dt-\sqrt{\frac{4\pi\xi^2}{\lambda_\text{th}^2}}dU\bigg\}.
		\end{split}
	\end{align}
    Considering the real and imaginary part of this equation and then Fourier transforming\footnote{As we are in an infinite homogeneous system the one particle basis are the plane waves.}
	\begin{align}
		\mathcal{F}\{\phi\}(\textbf{k})=\int d^n\textbf{r}e^{-i\textbf{k}\cdot\textbf{r}}\phi(\textbf{r})
	\end{align}
	we derive the equations (denoting $\mathcal{F}\left\{\Re (dW)\right\}=dW_1$, $\mathcal{F}\left\{\Im (dW)\right\}=dW_2$, $\mathcal{F}\left\{dU\right\}=\sqrt{\tilde{\varepsilon}_n(\textbf{k})}dW_3$)\footnote{In three dimensions this result was already derived in \cite{mcdonald_anomalous_2019}.}
	\begin{align}
		\label{linF}
		d\begin{bmatrix}
			\delta\tilde{n}(\textbf{k}) \\
			\tilde{\theta}(\textbf{k})
		\end{bmatrix}=-A(\textbf{k})\begin{bmatrix}
			\delta\tilde{n}(\textbf{k}) \\
			\tilde{\theta}(\textbf{k})
		\end{bmatrix}dt+B(\textbf{k})\begin{bmatrix}
			dW_1(\textbf{k})\\
			dW_2(\textbf{k})\\
			dW_3(\textbf{k})
		\end{bmatrix}.
	\end{align}
	Here
	\begin{align}
		\begin{split}
			A(\textbf{k})&=\begin{bmatrix}
				\gamma k^2/2+2\gamma & -k^2 \\
				1+k^2/4 & \gamma k^2/2+n_C\xi^{n}\tilde{\varepsilon}_n(\textbf{k})k^2
			\end{bmatrix},\\
			B(\textbf{k})&=\begin{bmatrix}
				0 & -\sqrt{\frac{4(2\pi)^{n+1}\gamma}{n_C\lambda_\text{th}^2\xi^{n-2}}} & 0 \\
				-\sqrt{\frac{(2\pi)^{n+1}\gamma}{n_C\lambda_\text{th}^2\xi^{n-2}}} & 0 & \sqrt{\frac{2(2\pi)^{n+1}\tilde{\varepsilon}_n(\textbf{k})}{\lambda_\text{th}^2\xi^{-2}}}
			\end{bmatrix},\\
			&\langle dW_i^*(\textbf{k},t)dW_j(\textbf{k}',t)\rangle=\delta_{ij}\delta(\textbf{k}-\textbf{k}')dt.
		\end{split}
	\end{align}
	Note that $\delta\tilde{n}(\textbf{k})=\delta\tilde{n}^*(-\textbf{k})$, $\delta\tilde{\theta}(\textbf{k})=\delta\tilde{\theta}^*(-\textbf{k})$ and $k<k_\text{cut}$.
	
	Having derived the linearized SPGPE in equation (\ref{linF}) we{ study its properties in the following sections}.
	
	\subsection{Equilibrium Population and Ideal Cut-off Choice}
	\label{population}
	{We calculate the variances of the density and phase fluctuations. To simplify notation we define the autocorrelations}
	\begin{align}
		\begin{split}
		G_{nn}(\textbf{k},t)&=\int \frac{d^n\textbf{k}'}{(2\pi)^n}\langle \delta \tilde{n}(\textbf{k},t)\delta \tilde{n}^*(\textbf{k}',0)\rangle,\\
		G_{n\theta}(\textbf{k},t)&=\int \frac{d^n\textbf{k}'}{(2\pi)^n}\langle \delta \tilde{n}(\textbf{k},t)\tilde{\theta}^*(\textbf{k}',0)\rangle,\\
		G_{\theta\theta}(\textbf{k},t)&=\int \frac{d^n\textbf{k}'}{(2\pi)^n}\langle \tilde{\theta}(\textbf{k},t)\tilde{\theta}^*(\textbf{k}',0)\rangle.
		\end{split}
	\end{align}
	{where the integration takes care of the delta-distribution appearing due to the limit of an infinite system (large box) considered here. We will also write $G_{ab}(\textbf{k})=G_{ab}(\textbf{k},0)$. The autocorrelations at $t=0$ can be calculated using the formula}~\cite{gardiner_stochastic_2009,mcdonald_anomalous_2019}
	\begin{align}
		\label{formula}
		\begin{split}
		&\begin{bmatrix}
			G_{nn}(\textbf{k})\ G_{n\theta}^*(\textbf{k})\\
			G_{n\theta}(\textbf{k})\ G_{\theta\theta}(\textbf{k})
		\end{bmatrix}\\
		&=\frac{\det(A) BB^T+[A-\text{tr}(A)]BB^T[A-\text{tr}(A)]^T}{2\text{tr}(A)\det(A)}.
		\end{split}
	\end{align}
	The density-phase correlations turn out to vanish in equilibrium while we obtain the density-density fluctuations
	\begin{align}
		\label{densdensfluc}
		G_{nn}(\textbf{k})=\frac{2\pi}{n_C\lambda_\text{th}^2\xi^{n-2}}\frac{1}{1+k^2/4}
	\end{align}
	and the phase-phase fluctuations
	\begin{align}
		\label{phasephasefluc}
		G_{\theta\theta}(\textbf{k})=\frac{2\pi}{n_C\lambda_\text{th}^2\xi^{n-2}}\frac{1}{k^2}.
	\end{align}
	{Here, the diveregence of phase fluctuations towards $k=0$ reflects that condensation in the continuum limit does only occur in three dimensions. A finite system size introduces an infrered cut-off and thereby resolves the divergence (the equations are only valid for $k>0$).}
	
	Therefore, the state with momentum $\textbf{k}$ has the single particle occupation (see appendix \ref{occupationcalculation})\footnote{{Strictly, in the SPGPE the occupation differs by $1/2$ to the value presented here due to the presence of virtual particles in the Wigner representation. For practical reasons it is desirable to not have to deal with these virtual particles and it is allowed to ignore them as long as the occupations in the coherent region are large against $1/2$}\cite{zurek_decoherence_1998}{. Fortunately, our analysis implies that this is indeed the case. In other words, we work in an approximation of the SPGPE focussing on thermal effects and ignoring quantum corrections. See }\ref{QuantumCorrection}{ for a short discussion of the occupation including the quantum corrections.}}
	\begin{align}
		\label{occupation}
		N_\textbf{k}=2\pi\frac{\xi^2}{\lambda_\text{th}^2}\frac{E_k}{\omega_k^2},
	\end{align}
	where $E_k=1+k^2/2$ is the energy in the free particle regime and $\omega_k=\sqrt{k^2(1+k^2/4)}$ is the dispersion relation. This result can also be derived by assuming the classical equipartition theorem and the virial theorem \cite{hadzibabic_two-dimensional_2011}. In particular, we have for small momenta in agreement with the Bogoliubov result (\ref{Bogoliubovoccsmallk})
	\begin{align}
		N_\textbf{k}\sim2\pi\frac{\xi^2}{\lambda_\text{th}^2}\frac{1}{k^2}=\frac{mk_\text{B}T}{p^2}.
	\end{align}
	\begin{figure}
		\centering
		\includegraphics[width=1\linewidth]{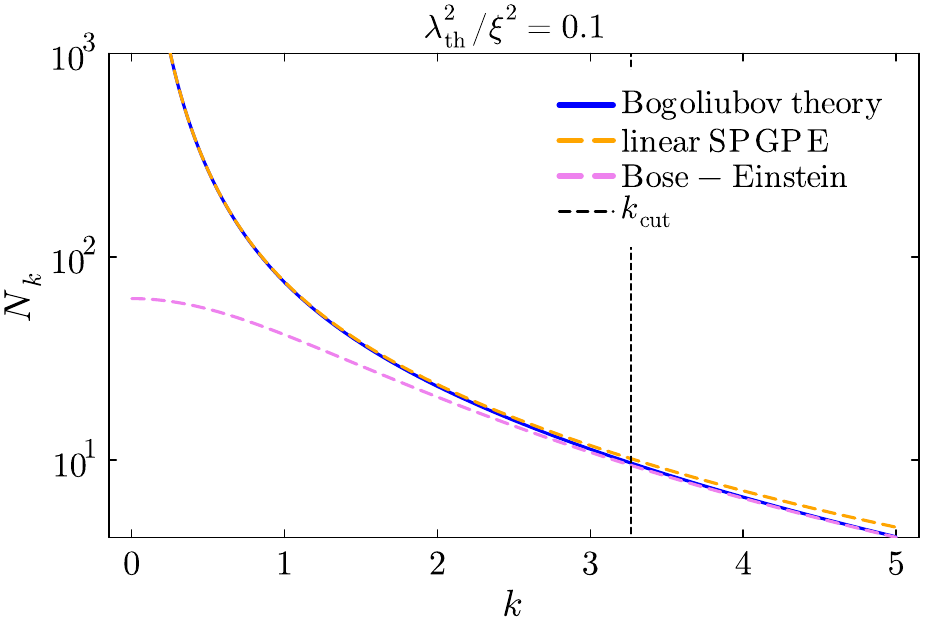}
		\caption{Occupation number according to the SPGPE (\ref{occupation}) (orange dashed line) and the Bose-Einstein distribution (\ref{thoccupation}) valid for high momenta (violet dashed line) used in its derivation compared to the Bogoliubov predictions (solid blue line). As expected, the SPGPE theory result agrees well for small momenta. The black dashed line highlights the momentum at which (\ref{cut-offchoice}) sets the cut-off: SPGPE treatment corresponds to an occupation number according to the orange dashed line to its left (for smaller $k$) and to the violet dashed line to its right (for larger $k$). {We choose $\lambda_\text{th}^2/\xi^2=0.1$ leading to a small step between the occupation according to the SPGPE and the Bose-Einstein statistics (roughly 8\%) at the cut-off $k_\text{cut}=3.26$ according to }(\ref{cut-offocc}). Smaller values of $\lambda_\text{th}^2/\xi^2$ reduce the size of this step, see figure \ref{fig:cut-off}(b).{ The Bogoliubov and linear SPGPE curves are valid for sufficently low temperature, so that Bogoliubov treatment is valid and hence $n_\text{tot}\simeq n_C\simeq n_0$. In a system of size $L$, only momenta fulfilling $k=2\pi/L$ are actually occupied.}}
		\label{fig:occupationnumber}
	\end{figure}
	Figure \ref{fig:occupationnumber} compares the occupation number predicted by the SPGPE found in equation (\ref{occupation}) with the occupation as predicted by the Bogoliubov theory (\ref{Bogoliubovocc}). It demonstrates good agreement for low momenta. The departure at high momenta signals where the cut-off should be chosen.
	
	The occupation of the ground state is given by
	\begin{align}
		N_0=n_C(L\xi)^n-\sum_{|\textbf{k}|<k_\text{cut}}N_\textbf{k},
	\end{align}
	where $L$ is the lateral length of the box in terms of healing length. In three dimensions and a large system size (\ref{occupation}) implies
	\begin{align}
		\begin{split}
			\sum_{|\textbf{k}|<k_\text{cut}}N_\textbf{k}&\simeq\left(\frac{L}{2\pi}\right)^3\int_{|\textbf{k}|<k_\text{cut}}d^3\textbf{k}N_\textbf{k}\\
			&=\frac{2L^3}{\pi}\frac{\xi^2}{\lambda_\text{th}^2}\left[k_\text{cut}-\text{atan}\left(\frac{k_\text{cut}}{2}\right)\right].
		\end{split}
	\end{align}
	The total particle number is then given by
	\begin{align}
		N_\text{tot}=n_C(L\xi)^n+\left(\frac{L}{2\pi}\right)^n\int_{|\textbf{k}|>k_\text{cut}}d^n\textbf{k}N_\textbf{k}^\text{th}.
	\end{align}
	
	The occupation number found in (\ref{occupation}) gives us a recipy on how to choose the cut-off. In deriving the SPGPE thermal atoms are assumed to be distributed according to (\ref{thoccupation}). Ideally, at the cut-off these two occupation numbers should coincide. In practice, this is only the case at infinite temperature (in using the SPGPE, we always make a finite error). However, we can demand that the occupation numbers should be as close as possible. More precisely, we choose the cut-off so that
	\begin{align}
		\label{cut-offchoice}
		\partial_k\frac{N_\textbf{k}}{N_\textbf{k}^\text{th}}\bigg|_{k=k_\text{cut}}=0.
	\end{align}
	{Choosing the cut-off according to }(\ref{cut-offchoice}){ minimises the difference between the Bose-Einstein distribution and the SPGPE, and is the main result of this section.} For a cut-off choosen larger than this, high energetic states in the coherent region are getting unphysically highly populated. On the other hand, a smaller cut-off choice leads to an underestimation of the particle number in the incoherent region and hence to an underestimation of the damping and noise strength.
	
	Approximating
	\begin{align}
		\begin{split}
			N_\textbf{k}^\text{th}&\simeq\frac{1}{\exp\left(\frac{\lambda_\text{th}^2}{2\pi\xi^2}\left[1+\frac{k^2}{2}\right]\right)-1}\\
			&\simeq\frac{1}{\frac{\lambda_\text{th}^2}{2\pi\xi^2}\left[1+\frac{k^2}{2}\right]+\left(\frac{\lambda_\text{th}^2}{2\pi\xi^2}\left[1+\frac{k^2}{2}\right]\right)^2}
		\end{split}
	\end{align}
	We find the momentum at the cut-off
	\begin{align}
		\label{kcut}
		\begin{split}
		k_\text{cut}^2&=2\left(4\pi\frac{\xi^2}{\lambda_\text{th}^2}+\sqrt{\left[4\pi\frac{\xi^2}{\lambda_\text{th}^2}\right]^2-1}\right)^{1/3}\\
		&\text{{$-2$}}+\frac{2}{\left(4\pi\frac{\xi^2}{\lambda_\text{th}^2}+\sqrt{\left[4\pi\frac{\xi^2}{\lambda_\text{th}^2}\right]^2-1}\right)^{1/3}}\\
		&\simeq 2\left[8\pi\frac{\xi^2}{\lambda_\text{th}^2}\right]^{1/3}-2.
		\end{split}
	\end{align}
	This implies the thermal occupation at the cut-off 
	\begin{align}
		\label{cut-offocc}
		N_\text{cut}\simeq\left[\pi\frac{\xi^2}{\lambda_\text{th}^2}\right]^{2/3}.
	\end{align}
	
	\begin{figure}
		\centering
		\includegraphics[width=1\linewidth]{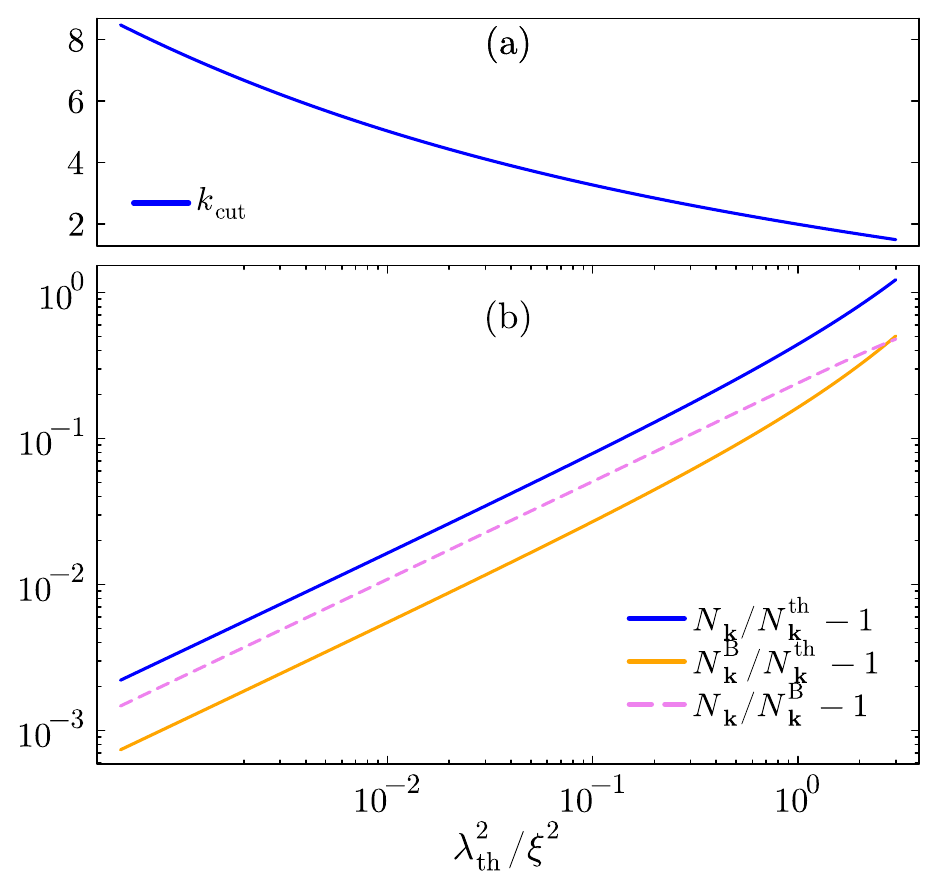}
		\caption{\textbf{(a)}: Cut-off choosen according to equation (\ref{cut-offchoice}). The normalized cut-off momentum $k_\text{cut}$ decreases with increasing $\lambda_\text{th}^2/\xi^2$. \textbf{(b)}: Particle number at the cut-off according to the linearized SPGPE (\ref{occupation}) compared to according to the Bose-Einstein statistics (blue line). While the difference for small $\lambda_\text{th}^2/\xi^2$ is negligible, for values $\lambda_\text{th}^2/\xi^2\gtrsim1$ the error becomes increasingly large and the use of the (high temperature) SPGPE invalid. We also show the deviation of the occupation according to the Bose-Einstein distribution and the linearized SPGPE to the occupation according to Bogoliubov theory (orange solid and violet dashed line, respectively).}
		\label{fig:cut-off}
	\end{figure}
	Figure \ref{fig:cut-off} shows the value of $k_\text{cut}$ chosen according to (\ref{cut-offchoice}) together with an estimate of the error made in the SPGPE given by the quotient of the occupation at the cut-off according to the linearized SPGPE and the Bose-Einstein statistics $N_{kcut}/N_{kcut}^\text{th}$. Validity of the SPGPE (in the sense of less than roughly twenty percent error in any occupation number) is demonstrated for values $\lambda_\text{th}^2/\xi^2\lesssim1$ verifying the argumentation in \cite{gardiner_stochastic_2003}\footnote{In \cite{gardiner_stochastic_2003} the requirement $k_\text{B}T/\mu\gtrsim10$ is given as a validity condition for SPGPE treatment. Since $\xi\simeq\hbar/\sqrt{m\mu}$ our analysis verifies that the SPGPE can be expected to become unreliable once $k_\text{B}T/\mu\lesssim10$.}. Use of the SPGPE is hence valid as long as the thermal de Broglie wavelength is shorter than a healing length\footnote{Note that only a small portion of the particles are assigned to the states with inaccurate occupation numbers, so that for many applications the SPGPE might still yield more accurate results as one may expect from the relative error in figure \ref{fig:cut-off}(b).}. Considering (\ref{cut-offocc}) implies that the particle number at the cut-off should hence always be larger than one.
	\subsection{Equilibration}
	\label{Decay}
	The previous section studied the equilibrium population according to the linearized SPGPE and thereby identified an optimized cut-off choice. This section is dedicated to the process of equilibration.
	
	We first note that the density and phase variances evolve according to (\ref{linF})
	\begin{align}
		\begin{split}
		\partial_t\langle|\delta\tilde{n}(\textbf{k})|^2\rangle&=-\gamma(4+k^2)\langle|\delta\tilde{n}(\textbf{k})|^2\rangle+k^2\langle\delta\tilde{n}\tilde{\theta}^*+c.c.\rangle\\
		&+4\frac{(2\pi)^{n+1}\gamma}{n_C\lambda_\text{th}^2\xi^{n-2}}\\
		\partial_t\langle|\tilde{\theta}(\textbf{k})|^2\rangle&=-(\gamma k^2+2n_C\xi^n\tilde{\varepsilon}_n(\textbf{k})k^2)\langle|\tilde{\theta}(\textbf{k})|^2\rangle\\
		&-(1+k^2/4)\langle\delta\tilde{n}\tilde{\theta}^*+c.c.\rangle\\
		&+\frac{(2\pi)^{n+1}\gamma}{n_C\lambda_\text{th}^2\xi^{n-2}}+\frac{2(2\pi)^{n+1}\tilde{\varepsilon}_n(\textbf{k})}{\lambda_\text{th}^2/\xi^{2}}\\
		\partial_t\langle\delta\tilde{n}\tilde{\theta}^*+c.c.\rangle&=-(4+k^2)/2\langle|\delta\tilde{n}(\textbf{k})|^2\rangle+2k^2\langle|\tilde{\theta}(\textbf{k})|^2\rangle\\
		&-[4\gamma+2\gamma k^2+2n_C\xi^n\tilde{\varepsilon}_n(\textbf{k})k^2]\langle\delta\tilde{n}\tilde{\theta}^*+c.c.\rangle
		\end{split}
	\end{align}
We can eliminate the density-phase correlations and either the phase-phase or the density-density correlations to obtain a third order differential equation for $\langle|\delta\tilde{n}(\textbf{k})|^2\rangle$, $\langle|\tilde{\theta}(\textbf{k})|^2\rangle$, respectively. As we can write the single-particle occupation number as {(see appendix }\ref{occupationcalculation})
\begin{align}
	\begin{split}
	\frac{N_\textbf{k}}{n_C\xi^n}&=\langle|\tilde{\psi}(\textbf{k})|^2\rangle\frac{d^n\textbf{k}}{(2\pi)^n}\\
	&\simeq\langle|\tilde{\theta}(\textbf{k})|^2\rangle\frac{d^n\textbf{k}}{(2\pi)^n}+\frac{\langle|\delta\tilde{n}(\textbf{k})|^2\rangle}{4}\frac{d^n\textbf{k}}{(2\pi)^n}
	\end{split}
\end{align}
we derive that it evolves according to (in lowest order of  $\Gamma_k$; $N_\textbf{k}^\text{eq}$ denotes the equilibrium population)
\begin{align}
	\label{occupationevol}
\partial_t^3N_\textbf{k}+8\Gamma_k\partial_t^2N_\textbf{k}+4\omega_k^2\partial_tN_\textbf{k}+8\omega_k^2\Gamma_kN_\textbf{k}=8\omega_k^2\Gamma_kN_\textbf{k}^\text{eq},
\end{align}
where we introduced the decay rate
\begin{align}
	\label{GammaknD}
	\Gamma_k&=\left(1+\frac{k^2}{2}\right)\gamma +\frac{n_C\xi^{n}}{2}\tilde{\varepsilon}_n(k)k^2.
\end{align}

Equation (\ref{occupationevol}) supports in lowest order of the damping the homogeneous solutions $\exp(-2\Gamma_kt)$, $\exp(-3\Gamma_kt)\cos(2\omega_kt)$ and $\exp(-3\Gamma_kt)\sin(2\omega_kt)$. If we start in an equilibrium state (density-phase correlations and their gradient vanish at $t=0$) the occupation number at time $t$ is in lowest order of the damping given by
\begin{align}
	\label{equilibration}
	N_\textbf{k}(t)=[N_\textbf{k}(0)-N_\textbf{k}^\text{eq}]e^{-2\Gamma_\textbf{k}t}+N_\textbf{k}^\text{eq}.
\end{align}
{This equation holds for $k>0$, while we conclude for the condensate occupation}
\begin{align}
	\label{N0growth}
	\text{{$N_0(t)=N_C(t)-\sum_{0<|\textbf{k}|<k_\text{cut}}N_\textbf{k}(t).$}}
\end{align}

\begin{figure}
	\centering
	\includegraphics[width=1\linewidth]{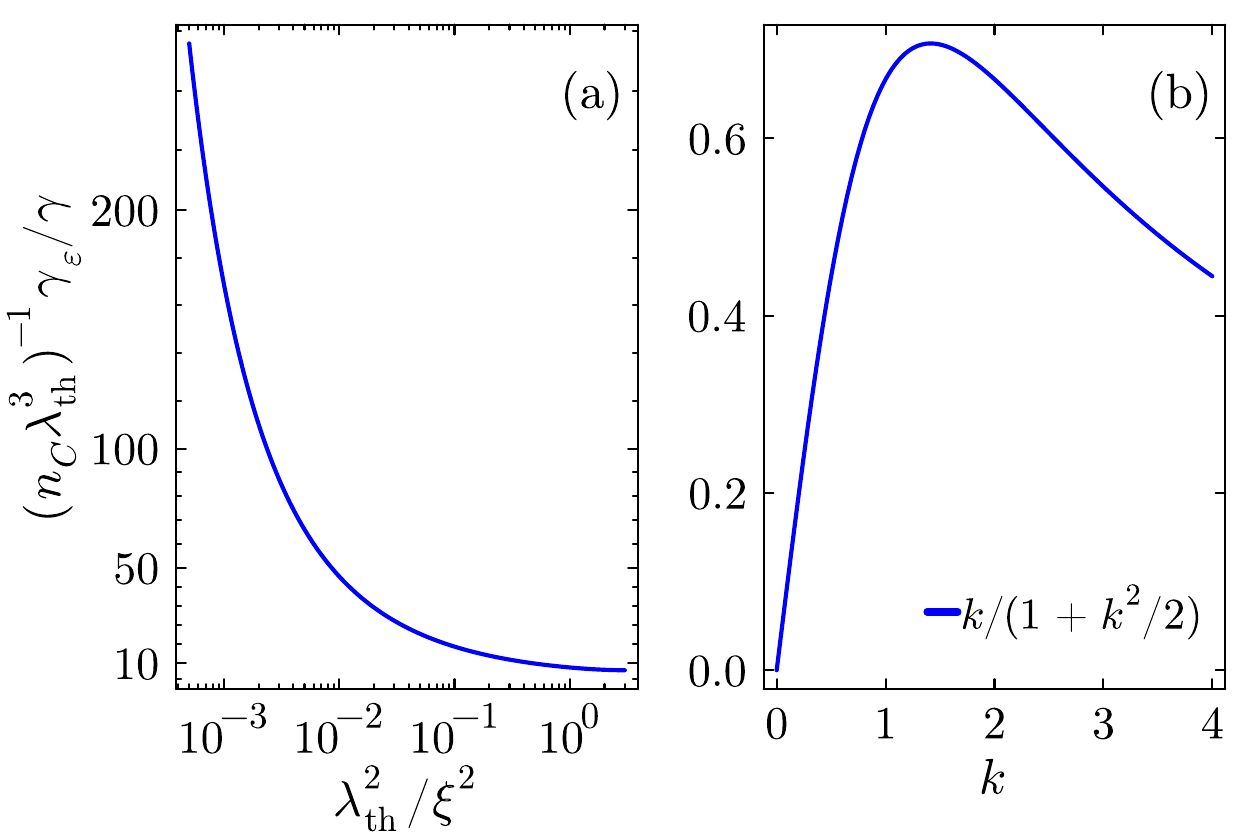}
	\caption{Comparison of the strength of number and energy damping. \textbf{(a)} shows the ratio $\gamma_\varepsilon/\gamma$ scaled by the phase space density $n_C\lambda_\text{th}^3$ for different values of $\lambda_\text{th}^2/\xi^2$. \textbf{(b)} displays the ratio between the scaling of the damping induced by energy and number damping with momentum $k${: according to} (\ref{Gammak}){ energy damping scales linear while number damping scales with $1+k^2/2$, so that their ratio scales with $k/(1+k^2/2)$}. The relative strength of energy damping to number damping in the decay rate $\Gamma_k$ corresponds to the product of the two graphs shown and the phase space density. Energy damping turns out to be the usually dominant process; only for $k\ll1$ does number damping dominate the decay.}
	\label{fig:energyvsnumber}
\end{figure}

We now have a closer look at the decay rate $\Gamma_k$. In the unreduced three dimensional case we have
\begin{align}
	\tilde{\varepsilon}_3(\textbf{k})=16\pi a_\mathrm{s}^2N_\text{cut}/(\xi^2k).
\end{align}
Defining
\begin{align}
	\gamma_\varepsilon=8\pi N_\text{cut}n_Ca_\text{s}^2\xi =\frac{1}{2\pi}\frac{N_\text{cut}}{n_C\xi^3}.
\end{align}
we can write
\begin{align}
	\label{Gammak}
	\Gamma_k=\gamma\left(1+\frac{k^2}{2}\right)+\gamma_\varepsilon k.
\end{align}
This allows us now to perform a direct comparison of the strength of number and energy damping. It is
\begin{align}
	\frac{\gamma_\varepsilon}{\gamma}=\pi n_C\lambda_\text{th}^2\xi\frac{N_\text{cut}}{\gamma\lambda_\text{th}^2/(8a_\text{s}^2)}.
\end{align}
Using the bound for $\gamma$ (and that $N_\text{cut}>1$) derived in (\ref{gambound}) we have
\begin{align}
	\frac{\gamma_\varepsilon}{\gamma}>n_C\lambda_\text{th}^3\frac{\xi}{\lambda_\text{th}}
\end{align}
As the validity of the SPGPE requires $\xi/\lambda_\text{th}>1$, energy damping dominates the decay close to and below the critical temperature, as then $n_C\lambda_\text{th}^3>1$.{ The linearization we are working in requires the slightly stronger condition $n_C\lambda_\text{th}^2\xi\gg1$ and we hence derive the main result of this section}
\begin{align}
	\text{{$\gamma_\varepsilon\gg\gamma.$}}
\end{align}
{See also figure }\ref{fig:energyvsnumber}(a).

As number damping scales quadratic with momentum and energy damping merely linear, naively number damping {always} dominates for large and small enough momenta. However, according to (\ref{Gammak}) in practice number damping only dominates for momenta larger
\begin{align}
	\begin{split}
	k_+&=\frac{\gamma_\varepsilon}{\gamma}+\sqrt{\left(\frac{\gamma_\varepsilon}{\gamma}\right)^2-2}\simeq2\frac{\gamma_\varepsilon}{\gamma},
	\end{split}
\end{align}
which usually lie outside the coherent region, or smaller than
\begin{align}
	\begin{split}
		k_-&=\frac{\gamma_\varepsilon}{\gamma}-\sqrt{\left(\frac{\gamma_\varepsilon}{\gamma}\right)^2-2}\simeq\frac{\gamma}{\gamma_\varepsilon}.
	\end{split}
\end{align}
Hence, number damping is only relevant in the relaxation of low lying states with momenta $k\leq\gamma/\gamma_\varepsilon\ll1$. As we will argue in the following, this holds notably \emph{not} for the condensate growth which is dominated by energy damping. Figure \ref{fig:energyvsnumber} displays the ratio between energy and number damping, again verifying the dominance of the former. 

The condensate occupation is given by
\begin{align}
	\begin{split}
	N_0&=N_C-\sum_{0<|\textbf{k}|<k_\text{cut}}N_\textbf{k}\\
	&\simeq N_C-\left(\frac{L}{2\pi}\right)^3\int_{|\textbf{k}|<k_\text{cut}}d^3\textbf{k}N_\textbf{k}.
	\end{split}
\end{align}
For an (unphysical\footnote{The use of the SPGPE requires the incoherent region to be in thermal equilibrium. However, the lowest lying states of the incoherent region will not equilibrate much faster than the highest lying states in the coherent region. Therefore, the temperature may not change much faster  with time than the decay of the latter $\partial_tT/T<2\Gamma_{k\text{cut}}$ for the SPGPE to be a valid description.}) instantaneous temperature quench from $T_0$ to $T$ the growth rate (or decline rate) of the condensate is hence given by
\begin{align}
	\begin{split}
	\partial_tN_0&=\partial_tN_C-\frac{2L^3}{\pi}\left(\frac{\xi^2}{\lambda_\text{th}^2(T)}-\frac{\xi^2}{\lambda_\text{th}^2(T_0)}\right)\\
	&\times\int_0^{k_\text{cut}}dk\frac{1+k^2/2}{1+k^2/4}\Gamma_ke^{-2\Gamma_kt}.
	\end{split}
\end{align}
For short times we have\footnote{We assume $n_C(T)\simeq n_C(T_0)$ (not to be confused with $n_0(T)$) valid at sufficiently low temperatures $T, T_0$. This also includes the neglection of $\partial_t N_C$, which is suppressed by $n_C\lambda_\text{th}^2\xi$ compared to the term in the second line\cite{mcdonald_dynamics_2020}.}
\begin{align}
	\begin{split}
		\partial_tN_0&=-\frac{2L^3}{\pi}\left[\frac{\xi^2}{\lambda_\text{th}^2(T)}-\frac{\xi^2}{\lambda_\text{th}^2(T_0)}\right]\\
		&\times\bigg\{\gamma\left[2k_\text{cut}-2\text{atan}\left(\frac{k_\text{cut}}{2}\right)\right]\\
		&+\gamma_\varepsilon\left[k_\text{cut}^2-2\log\left(1+\frac{k_\text{cut}^2}{4}\right)\right]\bigg\}.
		\end{split}
\end{align}
As $k_\text{cut}$ is of order one and we saw already earlier that $\gamma_\varepsilon\gg\gamma$, the early growth of the condensate after a quench is clearly dominated by energy damping.

A similar argumentation holds in the two and one dimensional case, in which energy damping can be expected to dominate for all states fulfilling $k>1/(n_{2\text{D}}\lambda_\text{th}^2)$ and $k>\xi/(n_{1\text{D}}\lambda_\text{th}^2)$, respectively. Again, the condensate growth will be dominated by energy damping, too\footnote{{In one and two dimensions the infrared divergence implies that excessively large systems have almost all of their (non condensate) particles at low momenta for which number damping dominates. However}, for realistic system sizes (up to hundreds of healing lengths) {most of the (non condensate) particles occupy higher lying states and} energy damping can be expected to dominate.}.

\begin{figure}
	\centering
	\includegraphics[width=1\linewidth]{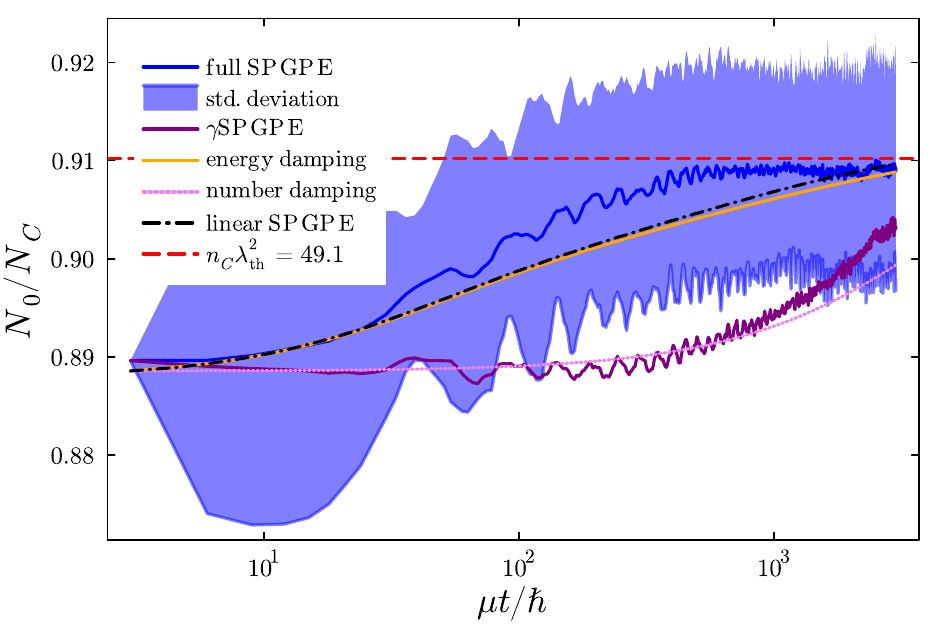}
	\caption{Condensate growth in a two dimensional BEC after a temperature quench from $n_C\lambda_\text{th}^2=38.9$ to $n_C\lambda_\text{th}^2=49.1$. The blue solid line shows a simulation of the full SPGPE ({5}00 trajectories) enveloped by its standard deviation (defined as $\sqrt{\langle N_0^2\rangle-\langle N_0\rangle^2}/\langle N_C\rangle$). Close agreement to (linear) energy damping (orange solid line{, calculated according to }(\ref{equilibration}), (\ref{N0growth}){ by ignoring the change in coherent particle number $N_C(t)\simeq N_C(0)$ and number damping and noise $\gamma=0$}) in the condensate growth is demonstrated. {The purple solid line shows a simulation of the $\gamma$SPGPE (ignoring energy damping and noise $\tilde{\varepsilon}=0$, 500 trajectories, same initial conditions as for the full SPGPE simulation), demonstrating that number damping can not explain the rapid condensate growth in the full SPGPE. The violet dotted line displays the prediction of the linearized $\gamma$SPGPE calculated according to }(\ref{equilibration}), (\ref{N0growth}){ by ignoring the change in coherent particle number $N_C(t)\simeq N_C(0)$ and energy damping and noise $\gamma=0$}). The red dashed line corresponds to the equilibrium state while the black dash-dotted line is the full linearized SPGPE result (again {employing }(\ref{equilibration}), (\ref{N0growth}){ and ignoring the change in coherent particle number $N_C(t)\simeq N_C(0)$})).}
	\label{fig:condensategrowth}
\end{figure}
Figure \ref{fig:condensategrowth} compares the linearized condensate growth in two dimensions to a simulation of the full SPGPE. Energy damping gives a good description of the growth. Notably, number damping plays only a minor role in the growth.
	
	\subsection{Fluctuations}
	\label{timedependence}
	In the previous section, we considered the temporal evolution of the occupation numbers. We now move on to characterize fluctuations in equilibrium. We calculate the autocorrelation and power spectrum.
	
	We first calculate the autocorrelation (see appendix \ref{FluctuationCalculation})
	\begin{align}
		\label{G}
		\begin{split}
			G(\textbf{k},t)&=\begin{bmatrix}
				G_{nn}(\textbf{k},t)\ G_{n\theta}^*(\textbf{k},t)\\
				G_{n\theta}(\textbf{k},t)\ G_{\theta\theta}(\textbf{k},t)
			\end{bmatrix}\\
			&\simeq\frac{2\pi}{n_C\lambda_\text{th}^2\xi^{n-2}}\exp\left(-\Gamma_k|t|\right)\\
			&\times\bigg(\cos(\Omega_kt)\begin{bmatrix}1/(1+k^2/4) & 0\\ 0 & 1/k^2\end{bmatrix}\\
			&+\frac{\sin(\Omega_kt)}{\Omega_k}\begin{bmatrix}
				0 & 1 \\
				-1 & 0
			\end{bmatrix}\bigg).
		\end{split}
	\end{align}
	We introduced the oscillation frequency
	\begin{align}
		\begin{split}
			\Omega_k&=\sqrt{k^2(1+k^2/4)-\left(\gamma-n_C\xi^{n}\tilde{\varepsilon}_n(\textbf{k})k^2/2\right)^2}. 
		\end{split}
	\end{align}
	
	Transforming into position space, we find the density-density autocorrelation ($|\textbf{r}'-\textbf{r}|=r$)
	\begin{align}
		\begin{split}
		&\langle \delta n(\textbf{r}',t)\delta n(\textbf{r},0)\rangle\\
		=&\begin{cases}
			\frac{2\xi}{n_C\lambda_\text{th}^2}\int_0^{k_\text{cut}}dk\frac{e^{-\Gamma_kt}}{1+k^2/4}\cos(\Omega_kt)\cos(kr),\ n=1\\
			\frac{1}{n_C\lambda_\text{th}^2}\int_0^{k_\text{cut}}dk\frac{ke^{-\Gamma_kt}}{1+k^2/4}\cos(\Omega_kt)J_0(kr),\ n=2\\
			\frac{1}{\pi n_C\lambda_\text{th}^2\xi}\int_0^{k_\text{cut}}dk\frac{k^2e^{-\Gamma_kt}}{1+k^2/4}\cos(\Omega_kt)\frac{\sin(kr)}{kr},\ n=3
		\end{cases}.
	\end{split}
	\end{align}
	We note in particular for $r=t=0$
	\begin{align}
		\label{densityvariance}
		\langle \delta n(\textbf{r})^2\rangle=\begin{cases}
			\frac{4\xi}{n_C\lambda_\text{th}^2}\text{atan}(k_\text{cut}/2),\ n=1\\
			\frac{2}{n_C\lambda_\text{th}^2}\ln\left(1+k_\text{cut}^2/4\right),\ n=2\\
			\frac{4}{\pi}\frac{1}{n_C\lambda_\text{th}^2\xi}\left[k_\text{cut}-2\text{atan}(k_\text{cut}/2)\right],\ n=3
		\end{cases}.
	\end{align}
	
	The phase-phase autcorrelation can be calculated by
	\begin{align}
		\begin{split}
			&\langle \theta(\textbf{r}',t)\theta(\textbf{r},0)\rangle\\
			=&\begin{cases}
				\frac{4\pi\xi}{n_C\lambda_\text{th}^2L}\sum_{0<k<k_\text{cut}}\frac{e^{-\Gamma_kt}}{k^2}\cos(\Omega_kt)\cos(kr),\ n=1\\
				\frac{2\pi}{n_C\lambda_\text{th}^2L^2}\sum_{0<|\textbf{k}|<k_\text{cut}}\frac{e^{-\Gamma_kt}}{k^2}\cos(\Omega_kt)J_0(kr),\ n=2\\
				\frac{1}{\pi n_C\lambda_\text{th}^2\xi}\int_0^{k_\text{cut}}dke^{-\Gamma_kt}\cos(\Omega_kt)\frac{\sin(kr)}{kr},\ n=3
			\end{cases},
		\end{split}
	\end{align}
	where in one and two dimensions we write the correlation as a sum as the {continuum} limit is ill defined. In three dimensions we have for $r\gg t$ (neglecting the damping and approximating $\Omega_k\simeq k$)
	\begin{align}
		\langle \theta(\textbf{r}',t)\theta(\textbf{r},0)\rangle=\frac{1}{\pi}\frac{1}{n_C\lambda_\text{th}^2\xi}\frac{\text{Si}(k_\text{cut}[r-t])+\text{Si}(k_\text{cut}[r+t])}{2r}.
	\end{align}
	{Here,}
	\begin{align}
		\text{{$\text{Si}(z)=\int_0^zdx\sin x/x$.}}
	\end{align}
	{denotes the sine integral.}
	
	For times $t\ll1/\Gamma_k$ the exponential decay can be neglected. Since typically $\Gamma_k\ll1$, in practice the precise nature of the damping and noise induced by the thermal cloud has no visible influence on the decay of the (spatial) correlations. Instead, their time-evolution is dominated by sound waves in agreement with measurements of phase coherence in one-dimensional boxes\cite{rauer_recurrences_2018}.
	
	Having derived the autocorrelation $G(\textbf{k},t)$, we can now perform a Fourier{ }transform in the time domain to obtain the power spectrum for phase and density fluctuations (valid for momenta large enough compared to the damping so that $\Omega_k\in\mathbb{R}$)
	\begin{align}
		\begin{split}
			&S(\textbf{k},\omega)\\
			&=\int dt G(\textbf{k},t)e^{-i\omega t}\\
			&=\frac{2\pi}{n_C\lambda_\text{th}^2\xi^{n-2}}\bigg\{\bigg(\frac{\Gamma_k}{\Gamma_k^2+(\omega-\Omega_k)^2}\\
			&+\frac{\Gamma_k}{\Gamma_k^2+(\omega+\Omega_k)^2}\bigg)\begin{bmatrix}1/(1+k^2/4) & 0\\ 0 & 1/k^2\end{bmatrix}\\
			&+\left(\frac{\Gamma_k/\Omega_k}{\Gamma_k^2+(\omega-\Omega_k)^2}-\frac{\Gamma_k/\Omega_k}{\Gamma_k^2+(\omega+\Omega_k)^2}\right)\begin{bmatrix} 0 & -i\\ i & 0\end{bmatrix}\bigg\}.
		\end{split}
	\end{align}
	We conclude the total power spectrum (see appendix \ref{occupationcalculation})
	\begin{align}
		\begin{split}
			S_\text{tot}(\textbf{k},\omega)&=\text{{$\int dt\int\frac{d^n\textbf{k}'}{(2\pi)^n} \langle\tilde{\psi}(\textbf{k},t)\tilde{\psi}^*(\textbf{k}',0)\rangle e^{-i\omega t}$}}\\
			&=\frac{2\pi}{n_C\lambda_\text{th}^2\xi^{n-2}}\bigg(\frac{\Gamma_k}{\Gamma_k^2+(\omega-\Omega_k)^2}\left[\frac{E_k}{\omega_k^2}+\frac{1}{\Omega_k}\right]\\
			&+\frac{\Gamma_k}{\Gamma_k^2+(\omega+\Omega_k)^2}\left[\frac{E_k}{\omega_k^2}-\frac{1}{\Omega_k}\right]\bigg)\\
			&\simeq\frac{2\pi}{n_C\lambda_\text{th}^2\xi^{n-2}}\bigg(\frac{\Gamma_k}{\Gamma_k^2+(\omega-\omega_k)^2}\frac{E_k+\omega_k}{\omega_k^2}\\
			&+\frac{\Gamma_k}{\Gamma_k^2+(\omega+\omega_k)^2}\frac{E_k-\omega_k}{\omega_k^2}\bigg).
		\end{split}
	\end{align}
	\begin{figure}
		\centering
		\includegraphics[width=1\linewidth]{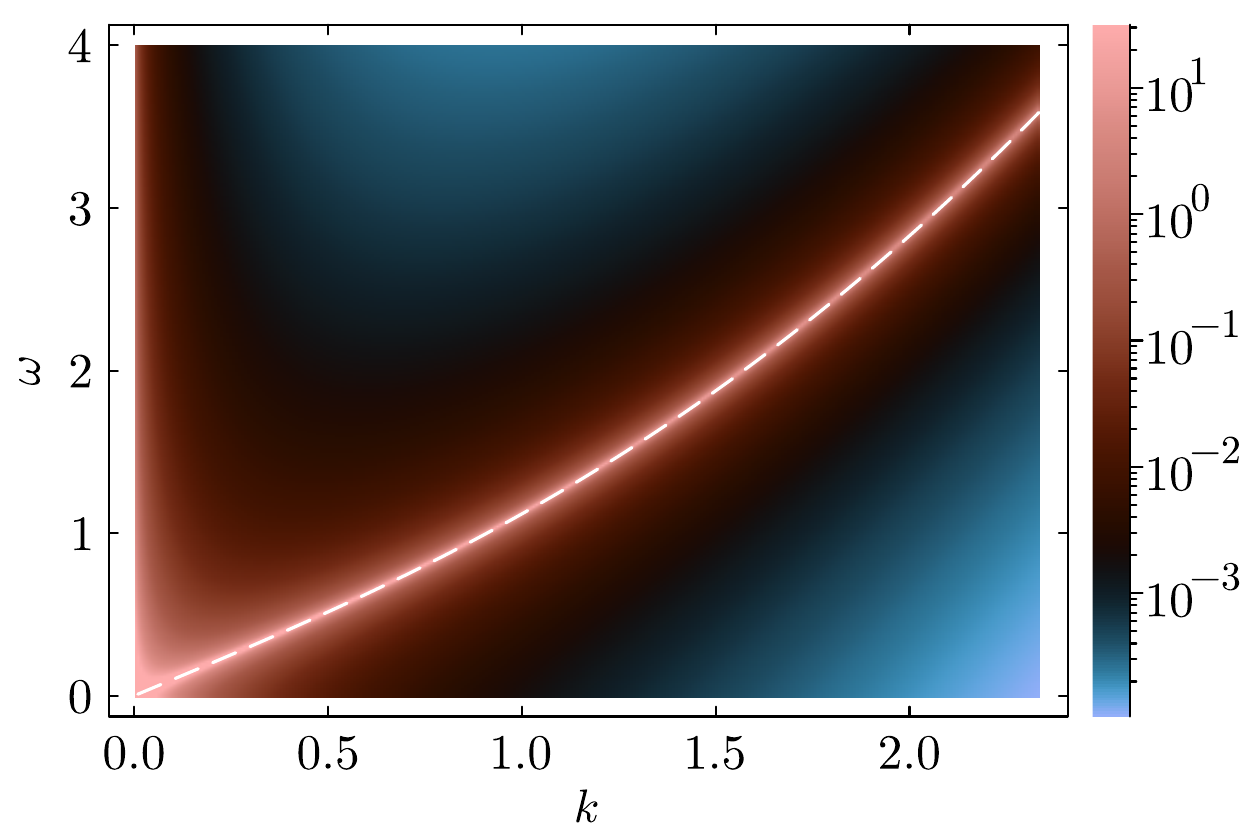}
		\caption{Total power spectrum $S_\text{tot}$ for a three dimensional Bose gas. $n_C\lambda_\text{th}^2\xi=50$, $\lambda_\text{th}^2/\xi^2=1/2$ and cut-off according to (\ref{cut-offchoice}). The white dashed line corresponds to the Bogoliubov dispersion around which the power spectrum is smeared out by the damping strength $\Gamma_k$ (\ref{Gammak}).}
		\label{fig:stot}
	\end{figure}
	The total power spectrum thus follows essentially the Bogoliubov dispersion, however smeared out by $\Gamma_k$ (see figure \ref{fig:stot}). As $\Gamma_k$ is dominated by energy damping for most $k$ energy damping determines the width.
	
	\subsection{Chemical Potential and Forward Scattering Hamiltonian}
	\label{Chemical}
	In the previous sections we implicitly assumed that $\mu= n_Cg$ and that $n_Ig=0$. However, strictly speaking this never holds. With the results found, we can calculate the next higher order correction. First, multiplying (\ref{SPGPE}) with $\psi^*$, integrating and averaging gives\cite{mcdonald_dynamics_2020} (contrary to the other parts of this paper the dimensionless SPGPE found in \ref{dimensionless} is not appropriate in this situation{; we nevertheless choose working in the dimensionless units as used so far in the following})
	\begin{align}
		\label{mueff}
		\begin{split}
		\frac{\mu_\text{eff}}{n_Cg}N_C&=\text{{$\int d^n\textbf{r}\frac{n_C\xi^n}{2}\langle|\nabla \psi|^2\rangle+n_C\xi^n\int d^n\textbf{r}\langle|\psi|^4\rangle$}}\\
		&\text{{$-L^n\int_{k<k_\text{cut}} \frac{d^n\textbf{k}}{(2\pi)^{n-1}}\frac{\xi^2}{\lambda_\text{th}^2}.$}}
		\end{split}
	\end{align}
	Here $\mu_\text{eff}=\mu-2n_Ig$ appreciates the forward scattering Hamiltonian and the term in the second line stems from the projection and corresponds to the single particle number of states multiplied by $k_\text{B}T$.
	
	Dividing by {$N_C$} gives (see appendix \ref{CPC})
	\begin{align}
		\label{muefflinear}
		\frac{\mu_\text{eff}}{n_Cg}\simeq1+\begin{cases}
			\frac{\xi}{n_C\lambda_\text{th}^2}\text{atan}(k_\text{cut}/2),\ n=1\\
			\frac{1}{n_C\lambda_\text{th}^2}\ln\left(1+k_\text{cut}^2/4\right),\ n=2\\
			\frac{2}{\pi}\frac{1}{n_C\lambda_\text{th}^2\xi}\left[k_\text{cut}-2\text{atan}(k_\text{cut}/2)\right],\ n=3
		\end{cases}.
	\end{align}
	If the scaled phase space density $n_C\lambda_\text{th}^2\xi^{n-2}$ becomes small, $\mu_\text{eff}\simeq n_Cg$ is hence not satisfied anymore and the argumentation of the previous sections becomes invalid.
	
	However, for states whose occupation is small compared to the particle number in the coherent region we {can argue as follows: first, we define}
	\begin{align}
		\phi(\textbf{r})\equiv\psi(\textbf{r})-\tilde{\psi}(\textbf{k}_0)\frac{e^{i\textbf{k}_0\cdot\textbf{r}}}{L^{n/2}}.
	\end{align}
	{In other words, we subtract the $\textbf{k}_0$ mode of $\psi$ from it. $\tilde{\phi}$ is then $\tilde{\psi}$, except at $\textbf{k}_0$ where we have $\tilde{\phi}(\textbf{k}_0)=0$. We} conclude for the interaction term
	\begin{align}
		\begin{split}
		|\psi(\textbf{r})|^2\psi(\textbf{r})&\simeq|\phi(\textbf{r})|^2\phi(\textbf{r})+2|\phi(\textbf{r})|^2\tilde{\psi}(\textbf{k}_0)\frac{e^{i\textbf{k}_0\cdot\textbf{r}}}{L^{n/2}}\\
		&+\phi(\textbf{r})^2\tilde{\psi}^*(\textbf{k}_0)\frac{e^{-i\textbf{k}_0\cdot\textbf{r}}}{L^{n/2}}.
		\end{split}
	\end{align}
	Performing a Fourier{ }transform {and considering $\textbf{k}=\textbf{k}_0$} we can neglect the first and the third term{: $|\phi|^2$ corresponds to real fluctuations around a constant value, while per construction $\tilde{\phi}(\textbf{k}_0)=0$, so that the first term vanishes. The third term becomes under the Fourier transform an integral over $\phi^2$, a fluctuating function with vanishing mean.} Additionally, {as we assumed a small occupation of the subtracted mode,} $|\phi|^2\simeq n_C$ under the Fourier transformation. In order to work with real equations we introduce the real and imaginary part of $\psi$ in terms of the real functions $\psi_r$, $\psi_i$ via
	\begin{align}
		\tilde{\psi}(\textbf{k})\equiv\tilde{\psi}_r(\textbf{k})+i\tilde{\psi}_i(\textbf{k}).
	\end{align}
	We deduce
	\begin{align}
		\label{linF2}
		d\begin{bmatrix}
			\tilde{\psi}_r(\textbf{k}) \\
			\tilde{\psi}_i(\textbf{k})
		\end{bmatrix}=-A(\textbf{k})\begin{bmatrix}
			\tilde{\psi}_r(\textbf{k}) \\
			\tilde{\psi}_\text{{$i$}}(\textbf{k})
		\end{bmatrix}dt+B(\textbf{k})\begin{bmatrix}
			dW_1(\textbf{k})\\
			dW_2(\textbf{k})
		\end{bmatrix},
	\end{align}
	where
	\begin{align}
		\begin{split}
			A(\textbf{k})&=\begin{bmatrix}
				\gamma\left(\frac{k^2}{2}+2-\frac{\mu_\text{eff}}{n_Cg}\right) & -\left(\frac{k^2}{2}+2-\frac{\mu_\text{eff}}{n_Cg}\right) \\
				\left(\frac{k^2}{2}+2-\frac{\mu_\text{eff}}{n_Cg}\right) & \gamma\left(\frac{k^2}{2}+2-\frac{\mu_\text{eff}}{n_Cg}\right)
			\end{bmatrix},\\
			B(\textbf{k})&=\begin{bmatrix}
				0 & -\sqrt{\frac{(2\pi)^{n+1}\gamma}{n_C\lambda_\text{th}^2\xi^{n-2}}}  \\
				-\sqrt{\frac{(2\pi)^{n+1}\gamma}{n_C\lambda_\text{th}^2\xi^{n-2}}} & 0 
			\end{bmatrix},\\
			&\langle dW_i^*(\textbf{k},t)dW_j(\textbf{k}',t)\rangle=\delta_{ij}\delta(\textbf{k}-\textbf{k}')dt.
		\end{split}
	\end{align}
	We neglected energy damping as we are interested in the equilibrium distribution which should not depend on the exact form of the damping\cite{gardiner_stochastic_2003}. However, for sufficiently high temperature we can expect number damping to dominate dynamics\footnote{We note that \emph{sufficiently high} will usually imply far above the critical temperature.}, so that in this case the dynamics are accurately described. We conclude {analogue to} (\ref{formula}) ($a,\ b\in{r,i}$)
	\begin{align}
		\begin{split}
		&\int \frac{d^n\textbf{k}'}{(2\pi)^n}\langle \tilde{\psi}_a(\textbf{k})\tilde{\psi}_b^*(\textbf{k}')\rangle\\ &=\frac{\pi}{n_C\lambda_\text{th}^2\xi^{n-2}}\frac{\delta_{ab}}{k^2/2+2-\mu_\text{eff}/(n_Cg)}
		\end{split}
	\end{align}
	and hence the occupation of high lying $\textbf{k}$ states
	\begin{align}
		\label{highkocc}
		N_\textbf{k}=2\pi\frac{\xi^2}{\lambda_\text{th}^2}\frac{1}{k^2/2+2-\mu_\text{eff}/(n_Cg)}.
	\end{align}
	
	We note that as soon as the occupation for high $\textbf{k}$ can be approximated by (\ref{highkocc}), an increase of $k_\text{cut}$ will likely worsen the accuracy of the SPGPE treatment: $N_\textbf{k}${/}$N_\textbf{k}^\text{th}$ diverge{s} with increasing $\textbf{k}$. A well chosen cut-off $k_\text{cut}$ should hence coincide with the onset of validity of (\ref{highkocc}) while $N_{k\text{cut}}/N_{k\text{cut}}^\text{th}-1$ will then give an estimate of the accuracy of the SPGPE.
	
	At high temperatures, we can expect all states to be only slightly populated and have uncorrelated phases. Hence, the argumentation should be valid for all $\textbf{k}$ and we can (implicitly) calculate the effective chemical potential at high temperatures (above the critical temperature). For convenience we introduce
	\begin{align}
		\bar{\mu}\equiv4-2\frac{\mu_\text{eff}}{n_Cg}=4\frac{n_\text{tot}}{n_C}-2\frac{\mu}{n_Cg}.
	\end{align}
	$\bar{\mu}$ can be calculated as the solution of the equation (see appendix \ref{CPC})
	\begin{align}
		\label{mueffnoco}
		\begin{split}
		\text{{$\frac{1}{\sqrt{\bar{\mu}}}\ \text{atan}\left(\frac{k_\text{cut}}{\sqrt{\bar{\mu}}}\right)$}}&\text{{$=\frac{n_C\lambda_\text{th}^2}{4\xi},\ n=1,$}}\\
		\text{{$\bar{\mu}$}}&\text{{$=\frac{k_\text{cut}^2}{e^{n_C\lambda_\text{th}^2}-1},\ n=2,$}}\\
		\text{{$\sqrt{\bar{\mu}}\ \text{atan}\left(\frac{k_\text{cut}}{\sqrt{\bar{\mu}}}\right)$}}&\text{{$=k_\text{cut}-\frac{\pi}{2}n_C\lambda_\text{th}^2\xi,\ n=3.$}}
		\end{split}
	\end{align}

\begin{figure}
	\centering
	\includegraphics[width=1\linewidth]{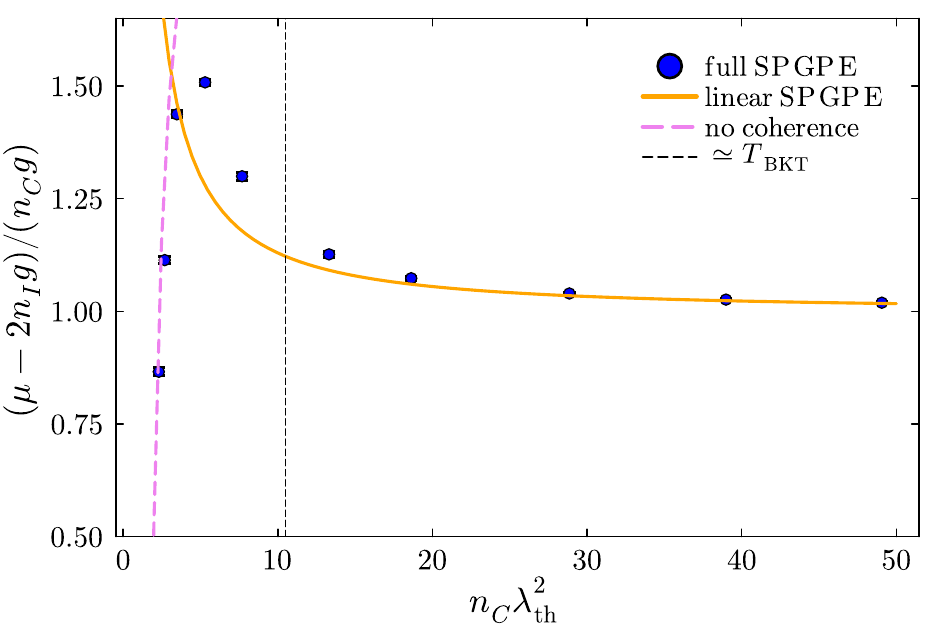}
	\caption{Effective chemical potential as a function of the phase space density in two spatial dimensions. The blue dots stem from simulations of the full SPGPE (with the cut-off chosen according to (\ref{cut-offchoice})), while the orange solid line corresponds to the analytical result in the linearized SPGPE (\ref{muefflinear}) valid for high phase space density. The violet dashed line corresponds to the limit of vanishing coherence (\ref{mueffnoco}), valid at low phase space density. For orientation we mark the position of the BKT transition as calculated from $n\lambda_\text{BKT}^2\simeq\ln(360n\xi^2)$ \cite{prokofev_critical_2001} by the vertical black dashed line.}
	\label{fig:muvsphasespace}
\end{figure}
Figure \ref{fig:muvsphasespace} compares the effective chemical potential in two dimensions in the two regimes of low and high phase space density calculated in this section with the result of simulations of the full SPGPE. It demonstrates good agreement of the results of this section with numerics away from the BKT transition.
	
	\section{Numerical Simulations}
	\label{Numeric}
	 \begin{figure*}
		\centering
		\includegraphics[width=1\linewidth]{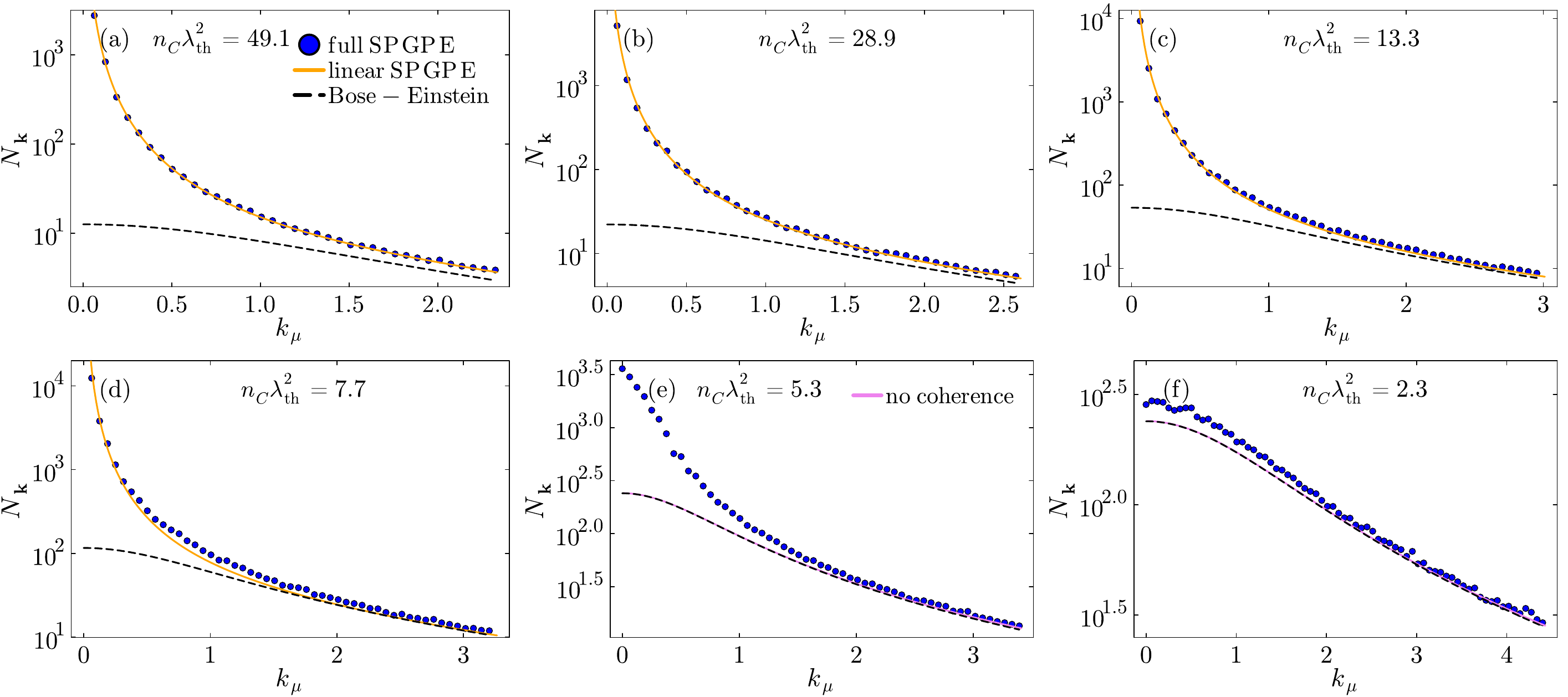}
		\caption{Comparison of the occupation numbers according to a simulation of the full (two-dimensional) SPGPE compared to the result of the linearized SPGPE (\ref{occupation}) ((a)-(d)) and the high temperature result (\ref{highkocc}) ((e) and (f)), respectively. We show additionally the distribution as expected by the Bose-Einstein statistics (\ref{2DBoseEinstein}) valid for high $k$, demonstrating increasing agreement to the SPGPE when approaching to the cut-off and increasing the temperature.{ The momentum along the $x$-axis is scaled by $\xi_\mu$: $k_\mu=\xi_\mu/\xi k$. The cut-off was chosen according to }(\ref{cut-offchoice}){ resulting in 2.3 (a), 2.6 (b), 3.0 (c), 3.3 (d), 3.4 (e) and 4.4 (f).} The deviation between the numerical and the analytical high temperature result in (e) and (f) implies a degree of coherence between low lying momentum states considerably above the BKT transition{ (which is reached for $n\lambda_\text{BKT}\simeq10.5$} \cite{prokofev_critical_2001}).}
		\label{fig:occupations}
	\end{figure*}
	We already demonstrated good agreement of our results in the low-temperature regime with the Bogoliubov theory and explored the validity of the (high temperature) SPGPE into this regime. To verify our analytical findings, we additionally run simulations of the full SPGPE as described in the following.
	
	We perform full SPGPE simulations using a weak semi-implicit Euler method akin to \cite{rooney_numerical_2014} on a $100\xi_{\mu}\times100\xi_{\mu}$-grid (where $\xi_\mu=\hbar/\sqrt{m\mu_\text{eff}}$)\footnote{In the simulations we define the healing length with regard to the effective chemical potential $\mu_\text{eff}$ as this is more convenient to implement numerically than the {definition} with regard to the (coherent region) density we used in the analytical parts. {As we are not working in the dimensionless form as in the rest of this work, we also note dimensionful quantities.}} with periodic boundary conditions and a time step between $0.01\hbar/\mu_\text{eff}$ and $0.001\hbar/\mu_\text{eff}$ (depending on the phase space density to achieve convergence of the implicit implementation of energy noise). We start with a homogeneous condensate and evolve it for more than $2\hbar/(\mu_\text{eff}\gamma)$ to reach equilibrium before extracting data for our analysis.
	
	We use the two-dimensional reduction as derived in \cite{bradley_low-dimensional_2015} and the integral kernel in (\ref{edampdef}) is hence set to\footnote{{Note that $\tilde{\varepsilon}_2$ is, as in the rest of the main part, scaled by $\xi^2$ compared to $\tilde{\varepsilon}$ in }(\ref{edampdef}).}
	\begin{align}
		\tilde{\varepsilon}_2(\textbf{k})=8\frac{a_\text{s}^2}{\text{{$\xi^2$}}}N_\text{cut}F\left(\left|\frac{l_z\textbf{k}}{2\xi}\right|^2\right),
	\end{align}
	where we assume a thickness of the atomic cloud of $l_z=\xi_\mu$. The energy cut-off is then (for $n_Cg\simeq\mu_\text{eff}$)
	\begin{align}
		\begin{split}
		\epsilon_\text{cut}&\simeq\frac{\hbar\omega_z}{2}+2\mu+\frac{\hbar^2k^2_\text{cut}}{2m\xi^2}\\
		&\simeq\left(\frac{5}{2}+\frac{k^2_\text{cut}}{2}\right)\mu_\text{eff}.
		\end{split}
	\end{align}
	We choose the momentum cut-off $k_\text{cut}$ (and thus the number of states) according to (\ref{cut-offchoice}). For all simulations we chose $n_C\xi^2=100$ allowing to study the SPGPE (which requires $\lambda_\text{th}^2/\xi^2<1$) for a large range of phase space densities $n_C\lambda_\text{th}^2$ in the presence of a considerable condensate fraction. We perform simulations for varying $n_C\lambda_\text{th}^2$ which then fix $\gamma$, $N_\text{cut}$ according to (\ref{gamdef}) and $N_\text{cut}=(\exp[\beta (\epsilon_\text{cut}-\mu_{3D})]-1)^{-1}$ (where $\mu_{3D}=\mu+\hbar\omega_z/2\simeq\mu_\text{eff}+\hbar\omega_z$) respectively\footnote{For initializing the parameters we employed the assumption $\mu_\text{eff}=n_Cg$ which leads to minor errors in the damping strengths for smaller $n_C\lambda_\text{th}^2$. However, this does not affect the equilibrium distribution and for larger $n_C\lambda_\text{th}^2$ the influence on the time evolution is also negligible. We calculate the actual $n_C\lambda_\text{th}^2$ by multiplying its initialized value by $\langle|\psi|^2\rangle$.}.
	
	Those particles in the thermal cloud that populate the ground state along $z$ are then assumed to be populated according to the Bose-Einstein statistics
	\begin{align}
		\label{2DBoseEinstein}
		N_\textbf{k}^\text{th}=\frac{1}{\exp\left(\frac{\lambda_\text{th}^2}{2\pi\xi^2}\left[2-\frac{\mu_\text{eff}}{n_Cg}+\frac{k^2}{2}\right]\right)-1},
	\end{align}
	where $\mu_\text{eff}=\mu_{3\text{D}}-\hbar\omega_z/2-2n_Ig$.
	 
	We generate an ensemble of equilibrium samples via a time evolution of $5000\hbar/\mu_\text{eff}$, employing ergodicity to study their statistical properties. We obtain the equilibrium distributions as shown in \ref{fig:occupations}. For high phase space density $n_C\lambda_\text{th}^2$ they demonstrate good agreement to the linearized result (\ref{occupation}). With decreasing phase space density this agreement worsens, especially for larger momenta, and the linearization breaks down close to the BKT transition. Further decreasing the phase space density leads to an approach to the high-temperature coherence free result (\ref{highkocc}). Note that the (high $k$) agreement of the coherent region with the Bose-Einstein distribution decreases with increasing phase space density, so that the validity of the SPGPE worsens. As we hold $n_C\xi^2$ constant, this is the expected behaviour.
	
	\section{Conclusions}
	\label{Discussion}
	This work presents an analysis of the linearized SPGPE, describing thermal dynamics in homogeneous BECs. In deriving the equilibrium distribution we identify an optimal cut-off choice, Eq.~(\ref{cut-offchoice}), balancing the error in the occupation of the coherent and incoherent states, thereby eliminating the only free parameter in the SPGPE. {We derive} a useful analytical expression for this cut-off, {Eq.~(}\ref{kcut}{)}, valid at high phase-space density and in box potentials. {We expect that the rule could also be straightforwardly extended} to arbitrarily small phase-space density and arbitrarily formed potentials (then requiring a numerical solution of the equilibrium population of single-particle states to be matched with the Bose-Einstein statistics at high energies).
	
	The derived optimal cut-off requires the thermal occupation at the cut-off to be {always} larger than one. In contrast, in \cite{pietraszewicz_classical_2015,pietraszewicz_complex_2018,pietraszewicz_classical_2018} the ideal cut-off choice in one dimension was investigated and a much higher cut-off was derived. However, the authors aimed at including the full kinetic energy and density fluctuations into the coherent region. The SPGPE treatment accepts a considerable portion of the kinetic energy to be stored in the thermal cloud and instead aims at a minimal error in the occupation of states. This implies a higher accuracy in the study of coherent region dynamics (e.g. condensate growth and the decay of non-linear structures).
	
	We derive an estimation of the error made in the use of the SPGPE, given by the quotient of the occupation number according to our linearized SPGPE results and Bose-Einstein statistics. {Our estimate verifies the }assumption in \cite{gardiner_stochastic_2003} that its validity requires the thermal de Broglie wavelength to be shorter than {the} healing length. Further, we calculate the (effective) chemical potential at high phase space density, allowing a full determination of all SPGPE parameters directly from experimental parameters, as discussed in appendix \ref{SPGPEParameters}.

    We also derive a phase autocorrelation function{. While its decay depends weakly on the damping, the correlations turn out to be essentially universal.} Dephasing is entirely dominated by sound waves. The power spectrum, however, is smeared out by {energy damping, demonstrating its importance. This is consistent with the single mode limit, where energy damping reduces to an exact phase diffusion}\cite{kordas_decay_2013}.
	
Studying the decay of an out-of-equilibrium state (such as after a quench) emphasizes the importance of the energy damping mechanism further. We find that it dominates the damping of most states and hence will dictate thermal evolution. Notably, the inclusion of energy damping is crucial in the description of the growth of the groundstate. {W}e work in a linearized regime and our results are hence only strictly valid far enough below the critical temperature{. It would be interesting to investigate if the dominance extends into the critical region}. Hence, our work points out the importance of a careful re-examination of previous SPGPE treatments in order to decide whether the SPGPE provides a full quantitative first-principles description of thermal dynamics in BECs.
	
{Our results are significant for quantitative open system theory of BECs and might motivate future quantitative experimental tests of finite-temperature effects. }While {they} are derived in a homogeneous monoatomic system, there is no reason to assume that either favors energy damping and it can be expected that the relevance of energy damping extends to various potential forms and atomic mixtures~\cite{bradley_stochastic_2014}.
	
\section*{Acknowledgements}
We are grateful to the Dodd-Walls Centre for Photonic and Quantum Technologies for financial support.
	
	\appendix
	
	\section{Calculation of the Occupation Number and Autocorrelation}
	\label{occupationcalculation}
	In this section we are calculating the occupation number in the high scaled phase space density regime $n_C\lambda_\text{th}^2\xi^{n-2}\gg1$. We can estimate the population of the single-particle states from
	\begin{align}
		\begin{split}
			\frac{N_\textbf{k}}{n_C\xi^n}&=\langle|\tilde{\psi}(\textbf{k})|^2\rangle\frac{d^n\textbf{k}}{(2\pi)^n}\\
			&=\int d^n\textbf{r}\int d^n\textbf{r}'e^{i\textbf{k}\cdot(\textbf{r}-\textbf{r}')}\\
			&\times\langle\sqrt{1+\delta n(\textbf{r})}\sqrt{1+\delta n(\textbf{r}')}e^{i[\theta(\textbf{r})-\theta(\textbf{r}')]}\rangle\frac{d^n\textbf{k}}{(2\pi)^n}\\
			&\simeq\int d^n\textbf{r}\int d^n\textbf{r}'e^{i\textbf{k}\cdot(\textbf{r}-\textbf{r}')}\left\langle1+\frac{\delta n(\textbf{r})\delta n(\textbf{r}')}{4}\right\rangle\\
			&\times\langle e^{i[\theta(\textbf{r})-\theta(\textbf{r}')]}\rangle\frac{d^n\textbf{k}}{(2\pi)^n}.\\
		\end{split}
	\end{align}
	Now assuming $\theta(\textbf{r})-\theta(\textbf{r}')$ to be Gaussian distributed we have
	\begin{align}
		\langle e^{i[\theta(\textbf{r})-\theta(\textbf{r}')]}\rangle=e^{-\langle[\theta(\textbf{r})-\theta(\textbf{r}')]^2\rangle/2}.
	\end{align}
	For the exponent we can calculate (employing (\ref{phasephasefluc}))
	\begin{align}
		\begin{split}
			&-\frac{\langle[\theta(\textbf{r})-\theta(\textbf{r}')]^2\rangle}{2}\\
			&=\langle\theta(\textbf{r})\theta(\textbf{r}')-\theta(0)^2\rangle\\
			&\simeq\int \frac{d^n\textbf{k}'}{(2\pi)^n}\frac{2\pi}{n_C\lambda_\text{th}^2\xi^{n-2}}\frac{e^{i\textbf{k}\cdot(\textbf{r}-\textbf{r}')}-1}{(k')^2}\Theta(k_\text{cut}-k')
		\end{split}
	\end{align}
	so that we conclude for large scaled phase space density $n_C\lambda_\text{th}^2\xi^{n-2}\gg1$
	\begin{align}
		\begin{split}
			&\langle e^{i[\theta(\textbf{r})-\theta(\textbf{r}')]}\rangle\\
			&\simeq1+\int \frac{d^n\textbf{k}'}{(2\pi)^n}\frac{2\pi}{n_C\lambda_\text{th}^2\xi^{n-2}}\frac{e^{i\textbf{k}'\cdot(\textbf{r}-\textbf{r}')}-1}{(k')^2}\Theta(k_\text{cut}-k').
		\end{split}
	\end{align}
	As
	\begin{align}
		\begin{split}
			&\int d^n\textbf{r}\int d^n\textbf{r}'e^{i\textbf{k}\cdot(\textbf{r}-\textbf{r}')}\int \frac{d^n\textbf{k}'}{(2\pi)^n}\\
			&\times\frac{2\pi}{n_C\lambda_\text{th}^2\xi^{n-2}}\frac{e^{i\textbf{k}'\cdot(\textbf{r}-\textbf{r}')}-1}{(k')^2}\Theta(k_\text{cut}-k')\frac{d^n\textbf{k}}{(2\pi)^n}\\
			&=\frac{2\pi}{n_C\lambda_\text{th}^2\xi^{n-2}}\frac{\Theta(k_\text{cut}-k)}{k^2}
		\end{split}
	\end{align}
	and we have from (\ref{densdensfluc})
	\begin{align}
		\begin{split}
			&\int d^n\textbf{r}\int d^n\textbf{r}'e^{i\textbf{k}\cdot(\textbf{r}-\textbf{r}')}\left\langle\frac{\delta n(\textbf{r})\delta n(\textbf{r}')}{4}\right\rangle\frac{d^n\textbf{k}}{(2\pi)^n}\\
			&\simeq\frac{2\pi}{n_C\lambda_\text{th}^2\xi^{n-2}}\frac{\Theta(k_\text{cut}-k)}{4+k^2}
		\end{split}
	\end{align}
	the occupation number of the state with momentum $\textbf{k}\neq0$ is given by
	\begin{align}
		\begin{split}
			N_\textbf{k}&=2\pi\frac{\xi^2}{\lambda_\text{th}^2}\frac{E_k}{\omega_k^2}+\mathcal{O}\left(\frac{1}{n_C\lambda_\text{th}^2\xi^{n-2}}\frac{\xi^2}{\lambda_\text{th}^2}\right),
		\end{split}
	\end{align}
	where $E_k=1+k^2/2$ is the energy in the free particle regime and $\omega_k=\sqrt{k^2(1+k^2/4)}$ is the dispersion relation.
	
	 A {similar} argumentation holds for the autocorrelation{. Namely, it is (assuming that $\theta(\textbf{r},t)-\theta(\textbf{r}',0)$ is Gaussian distributed)}
	 \begin{align}
	 	\begin{split}
	 	\langle\delta &n(\textbf{r},t)e^{i[\theta(\textbf{r},t)-\theta(\textbf{r}',0)]}\rangle\\
	 	&=\sum_j\frac{1}{j!}\langle\delta n(\textbf{r},t)i^j[\theta(\textbf{r},t)-\theta(\textbf{r}',0)]^j\rangle\\
	 	&=\sum_{j\text{ odd}}\frac{j!!}{j!}i^j\langle\delta n(\textbf{r},t)[\theta(\textbf{r},t)-\theta(\textbf{r}',0)]\rangle\\
	 	&\times\left(\langle[\theta(\textbf{r},t)-\theta(\textbf{r}',0)]^2\rangle\right)^\frac{j-1}{2}\\
	 	&\simeq i\langle\delta n(\textbf{r},t)[\theta(\textbf{r},t)-\theta(\textbf{r}',0)]\rangle,
	 	\end{split}
	 \end{align}
	 {where $!!$ denotes the double factorial. The last approximation is justified by the inverse scaling of the autocorrelations with the scaled phase space density (see equation }(\ref{G}){ (note, however, that in 1D and 2D this argument is only valid for finite system sizes). We conclude}
	 \begin{align}
	 	\begin{split}
	 	&\langle\tilde{\psi}(\textbf{k},t)\tilde{\psi}^*(\textbf{k}',0)\rangle\frac{d^n\textbf{k}'}{(2\pi)^n}\\
	 	&\simeq\bigg(\langle\tilde{\theta}(\textbf{k},t)\tilde{\theta}^*(\textbf{k}',0)\rangle+\frac{1}{4}\langle\delta\tilde{ n}(\textbf{k},t)\delta\tilde{ n}^*(\textbf{k}',0)\rangle\\
	 	&+\frac{i}{2}\langle\tilde{\theta}(\textbf{k},t)\delta\tilde{n}^*(\textbf{k}',0)\rangle-\frac{i}{2}\langle\delta\tilde{n}(\textbf{k},t)\tilde{\theta}^*(\textbf{k}',0)\rangle\bigg)\frac{d^n\textbf{k}'}{(2\pi)^n}
	 	\end{split}
	 \end{align}
 and hence the result for the total power spectrum $S_\text{tot}$ in the main text.
 
 \section{Quantum corrections in the SPGPE}
 \label{QuantumCorrection}
 {Throughout the main part, we handled the SPGPE fluctuations as solely stemming from thermal effects. However, if interpreted more carefully, it does contain some order of quantum corrections: The SPGPE is derived by mapping a Wigner function to a stochastic equation}\cite{gardiner_stochastic_2002,gardiner_stochastic_2003}{. The resulting classical field keeps half a particle vacuum occupation per mode from the Wigner function. The real particle occupation of the state with momentum $\textbf{k}$ in a proper treatment of the SPGPE is hence one half lower than stated in} (\ref{occupation}):
 \begin{align}
 	\label{occupationcorrect}
 	N_\textbf{k}=2\pi\frac{\xi^2}{\lambda_\text{th}^2}\frac{E_k}{\omega_k^2}-\frac{1}{2}.
 \end{align}
{It thus captures some of the quantum depletion in the Bogoliubov result }(\ref{Bogoliubovocc})
\begin{align}
	\begin{split}
		N_\textbf{k}^B&=\frac{E_k/\omega_k}{\exp\left(\frac{\lambda_\text{th}^2}{2\pi\xi^2}\omega_k\right)-1}+\frac{1}{2}\left[\frac{E_k}{\omega_k}-1\right]\\
		&\simeq\frac{E_k/\omega_k}{\frac{\lambda_\text{th}^2}{2\pi\xi^2}\omega_k+\left(\frac{\lambda_\text{th}^2}{2\pi\xi^2}\omega_k\right)^2/2}+\frac{1}{2}\left[\frac{E_k}{\omega_k}-1\right]+\mathcal{O}\left(\frac{\lambda_\text{th}^2}{\xi^2}\right)\\
		&\simeq 2\pi\frac{\xi^2}{\lambda_\text{th}^2}\frac{E_k}{\omega_k^2}-\frac{1}{2}+\mathcal{O}\left(\frac{\lambda_\text{th}^2}{\xi^2}\right).
	\end{split}
\end{align}

{In the main part we work in a thermal approximation of the SPGPE neglecting these quantum corrections for practical purposes (e.g. their inclusion implies that we can not interpret $|\psi|^2$ as the density of coherent band atoms). Including the quantum corrections becomes important once $\lambda_\text{th}\simeq\xi$. As the proper SPGPE occupation }(\ref{occupationcorrect}){ and the Bose-Einstein distribution intersect, we can expect an optimal cut-off at this intersection. $k_\text{cut}$ then becomes the solution of}
\begin{align}
	\label{cut-offchoice2}
	2\pi\frac{\xi^2}{\lambda_\text{th}^2}\frac{E_k}{\omega_k^2}-\frac{1}{2}=\frac{1}{\exp\left(\frac{\lambda_\text{th}^2}{2\pi\xi^2}\left[1+\frac{k^2}{2}\right]\right)-1}.
\end{align}

\begin{figure}
	\centering
	\includegraphics[width=1\linewidth]{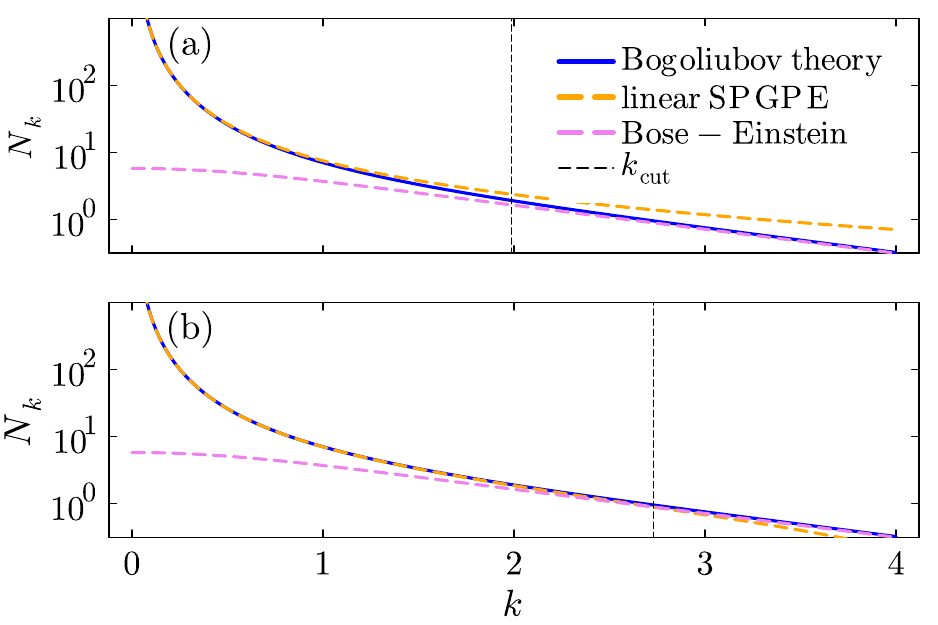}
	\caption{Comparison between the linear SPGPE occupation number \textbf{(a)} not appreciating (\ref{occupation}) and \textbf{(b)} appreciating (\ref{occupationcorrect}) the half particle virtual occupation. We also show the Bogoliubov prediction (\ref{Bogoliubovocc}) and the Bose-Einstein distribution (\ref{thoccupation}) as well as the cut-off choosen according to \textbf{(a)} (\ref{cut-offchoice}) and \textbf{(b)} (\ref{cut-offchoice2}), respectively. We choose $\lambda_\text{th}^2/\xi^2=1$ resulting in a considerable improvement of the predicted occupation if acknowledging the presence of virtual particles.}
	\label{fig:figure8}
\end{figure}

{Figure }\ref{fig:figure8}{ compares the occupation with and without appreciating the virtual particles with the Bogoliubov result and the Bose-Einstein distribution at $\lambda_\text{th}^2/\xi^2=1$. The appreciation of the virtual particles greatly improves the agreement of the occupations in this regime and suggests a larger optimal cut-off. As long as $\lambda_\text{th}\ll\xi$, however, the occupation of all states in the coherent region is large and the error made by ignoring the vacuum occupation is negligible}\cite{zurek_decoherence_1998}.

 \section{Fluctuation calculations}
 \label{FluctuationCalculation}
 In this section we present calculations performed for section \ref{timedependence}.{ We employ the formula}~\cite{gardiner_stochastic_2009}
 \begin{align}
 	G(\textbf{k},t)=e^{-At}G(\textbf{k}).
 \end{align}
 
 We first calculate using that $A$ is a $2\times2$-matrix
 \begin{align}
 	\begin{split}
 		e^{-At}&= \exp\left(-\Gamma_kt\right)\bigg(\cos(\Omega_kt)-\frac{\sin(\Omega_kt)}{\Omega_k}\\
 		&\times\begin{bmatrix}
 			\gamma-n_C\xi^{n}\tilde{\varepsilon}_n(\textbf{k})k^2/2 & -k^2 \\
 			1+k^2/4 & -\gamma+n_C\xi^{n}\tilde{\varepsilon}_n(\textbf{k})k^2/2
 		\end{bmatrix}\bigg),
 	\end{split}
 \end{align}
 where we introduced the oscillation frequency
 \begin{align}
 	\begin{split}
 		\Omega_k&=\sqrt{k^2(1+k^2/4)-\left(\gamma-n_C\xi^{n}\tilde{\varepsilon}_n(\textbf{k})k^2/2\right)^2}. 
 	\end{split}
 \end{align}
 We conclude the autocorrelation
 \begin{align}
 	\begin{split}
 		&G(\textbf{k},t)\\
 		&=\frac{2\pi}{n_C\lambda_\text{th}^2\xi^{n-2}}\exp\left(-\Gamma_k|t|\right)\\
 		&\times\bigg(\cos(\Omega_kt)\begin{bmatrix}1/(1+k^2/4) & 0\\ 0 & 1/k^2\end{bmatrix}\\
 		&-\frac{\sin(\Omega_k|t|)}{\Omega_k}\begin{bmatrix}
 			\frac{\gamma}{1+k^2/4}-\frac{n_C\xi^{n-2}\tilde{\varepsilon}_n(\textbf{k})k^2}{2+k^2/2} & -\text{sgn}(t) \\
 			\text{sgn}(t) & -\frac{\gamma}{k^2}+\frac{n_C\xi^{n}\tilde{\varepsilon}_n(\textbf{k})}{2}
 		\end{bmatrix}\bigg).
 	\end{split}
 \end{align}
 
 As the damping is small ($\gamma, n_C\xi^{n}\tilde{\varepsilon}_n(\textbf{k})\ll1$), we can ignore it inside the brackets and derive that the autocorrelation oscillates between diagonal and off-diagonal terms with a frequency determined by the dispersion relation
 \begin{align}
 	\Omega_k\simeq\omega_k=\sqrt{k^2(1+k^2/4)}.
 \end{align}
 We have the approximated autocorrelation
 \begin{align}
 	\begin{split}
 		G(\textbf{k},t)&\simeq\frac{2\pi}{n_C\lambda_\text{th}^2\xi^{n-2}}\exp\left(-\Gamma_k|t|\right)\\
 		&\times\bigg(\cos(\Omega_kt)\begin{bmatrix}1/(1+k^2/4) & 0\\ 0 & 1/k^2\end{bmatrix}\\
 		&+\frac{\sin(\Omega_kt)}{\Omega_k}\begin{bmatrix}
 			0 & 1 \\
 			-1 & 0
 		\end{bmatrix}\bigg).
 	\end{split}
 \end{align}
	 
	 \section{Chemical Potential Calculations}
	 \label{CPC}
	 Here we consider the different terms on the right hand side of (\ref{mueff}). We start with the high phase space density case. First, the interaction energy gives {(note that in the dimensionless units employed enegies are scaled by $n_Cg$)}
	 \begin{align}
	 	\text{{$n_C\xi^n\int d^n\textbf{r}\langle|\psi|^4\rangle=N_C+L^n\langle\delta n^2\rangle n_C\xi^n,$}}
	 \end{align}
	 where we calculated $\langle\delta n^2\rangle$ in (\ref{densityvariance}).
	 
	 The kinetic energy gives
	 \begin{align}
	 	\begin{split}
	 		&\left\langle\int d^n\textbf{r}\text{{$\frac{n_C\xi^n}{2}$}}|\nabla\psi|^2\right\rangle\\
	 		&=\text{{$\frac{n_C\xi^n}{2}$}}\int \frac{d^n\textbf{k}}{(2\pi)^n}\textbf{k}^2\langle|\tilde{\psi}(\textbf{k})|^2\rangle\\
	 		&\simeq\text{{$\frac{1}{2}$}}L^n\int \frac{d^n\textbf{k}}{(2\pi)^n}\textbf{k}^2N_\textbf{k}\\
	 		&=\begin{cases}
	 			\text{{$2\frac{\xi^2}{\lambda_\text{th}^2}L$}}\left[k_\text{cut}-\text{atan}(k_\text{cut}/2)\right],\ n=1\\
	 			\text{{$\frac{1}{2}\frac{\xi^2}{\lambda_\text{th}^2}L$}}\left[k_\text{cut}^2-2\ln\left(1+k_\text{cut}^2/4\right)\right],\ n=2\\
	 			\text{{$\frac{1}{\pi}\frac{\xi^2}{\lambda_\text{th}^2}L^3$}}\left[k_\text{cut}^3/3-2k_\text{cut}+4\text{atan}(k_\text{cut}/2)\right],\ n=3
	 		\end{cases}.
	 	\end{split}
	 \end{align}
	 
	 Lastly, the number of single particle states gives
	 \begin{align}
	 	\text{{$L^n\int_{k<k_\text{cut}} \frac{d^n\textbf{k}}{(2\pi)^{n-1}}\frac{\xi^2}{\lambda_\text{th}^2}$}}=\text{{$\frac{\xi^2}{\lambda_\text{th}^2}$}}\begin{cases}
	 		2Lk_\text{cut},\ n=1\\
	 		L^2k_\text{cut}^2/2\ n=2\\
	 		L^3k_\text{cut}^3/(3\pi),\ n=3
	 	\end{cases}.
	 \end{align}
	 
	 We now consider the high temperature case. First, $\psi$ becomes essentially Gaussian and hence
	 \begin{align}
	 	\langle|\psi|^4\rangle=2\langle|\psi|^2\rangle^2+\langle\psi^*\psi^*\rangle\langle\psi\psi\rangle\simeq2.
	 \end{align}
	 Thus, the interaction energy gives
	 \begin{align}
	 	\text{{$n_C\xi^n\int d^2\textbf{r}\langle|\psi|^4\rangle=2N_C$}}.
	 \end{align}
	 
	 For the kinetic energy we find
	\begin{align}
		\begin{split}
			&\left\langle\int d^2\textbf{r}\text{{$\frac{n_C\xi^n}{2}$}}\nabla\psi|^2\right\rangle\\
			&\simeq\text{{$\frac{1}{2}$}}L^n\int \frac{d^n\textbf{k}}{(2\pi)^n}\textbf{k}^2N_\textbf{k}\\
			&=\begin{cases}
				\begin{split}
				&\text{{$2\frac{\xi^2}{\lambda_\text{th}^2}L$}}\bigg[k_\text{cut}-\sqrt{2\{2-\mu_\text{eff}/(n_Cg)\}}\\
				&\times\text{atan}\left(\frac{k_\text{cut}}{\sqrt{2\{2-\mu_\text{eff}/(n_Cg)\}}}\right)\bigg]
				\end{split},\ n=1\\
				\begin{split}
				&\text{{$\frac{1}{2}\frac{\xi^2}{\lambda_\text{th}^2}L^2$}}\bigg[k_\text{cut}^2-2\{2-\mu_\text{eff}/(n_Cg)\}\\
				&\times\ln\left(1+k_\text{cut}^2/\{4-2\mu_\text{eff}/(n_Cg)\}\right)\bigg]
				\end{split},\ n=2\\
				\begin{split}
				&\text{{$\frac{1}{\pi}\frac{\xi^2}{\lambda_\text{th}^2}L^3$}}\bigg[k_\text{cut}^3/3-2\left\{2-\frac{\mu_\text{eff}}{n_Cg}\right\}\\
				&\times k_\text{cut}+\left(2\left\{2-\frac{\mu_\text{eff}}{n_Cg}\right\}\right)^{3/2}\\
				&\times\text{atan}\left(\frac{k_\text{cut}}{\sqrt{2\{2-\mu_\text{eff}/(n_Cg)\}}}\right)\bigg]
				\end{split},\ n=3
			\end{cases}.
		\end{split}
	\end{align}
	
	{Substituting into equation }(\ref{mueff}) we hence deduce the implicit equation given in (\ref{mueffnoco}).
	
	\section{SPGPE Parameters in a 3D box}
	\label{SPGPEParameters}
	We identified an ideal cut-off choice and calculated the effective chemical potential $\mu_\text{eff}/(n_Cg)$ in equilibrium. These results fix all parameters in the SPGPE. Here, we collect how to determine the parameters in the three dimensional dimensionless SPGPE $\lambda_\text{th}^2/\xi^2$ and $n_C\lambda_\text{th}^2\xi$ given the interaction strength $g$, the particle mass $m$, the system size {$L_d$}\footnote{Beware that {$L_d=L\xi$ in this section is the dimensionful system size in opposition to the dimensionless system size $L$ considered in the main part.We use, however, the scaled momentum $k$ from the main part}.}, the total particle number $N_\text{tot}$ and either the temperature $T$ or the ground state occupation $N_0$.
	
	In three dimensions the equation
	\begin{align}
		N_\text{tot}=N_C+\frac{\text{{$L_d$}}^3}{2\pi^2\xi^3}\int_{k_\text{cut}}^\infty dk k^2N_k^\text{th}
	\end{align}
	has to be solved for $N_C$. The expressions appearing on the right hand side can be written as (for high enough scaled phase space density and employing a cut-off choice according to (\ref{kcut}))
	\begin{align}
		\begin{split}
			N_\textbf{k}^\text{th}&=\frac{1}{\exp\left(\frac{\lambda_\text{th}^2}{2\pi\xi^2}\left[2-\frac{\mu_\text{eff}}{n_Cg}+\frac{k^2}{2}\right]\right)-1},\\
			k_\text{cut}&=2\left[8\pi\frac{\xi^2}{\lambda_\text{th}^2}\right]^{1/3}-2,\\
			\xi^3&=\left[\frac{\hbar^2 \text{{$L_d$}}^3}{mN_Cg}\right]^{3/2},\\
			\frac{\lambda_\text{th}^2}{\xi^2}&=2\pi\frac{N_Cg}{\text{{$L_d$}}^3k_\text{B}T},\\
			\frac{\mu_\text{eff}}{n_Cg}&=1+
				\frac{2}{\pi}\frac{\text{{$L_d$}}^3}{N_C\lambda_\text{th}^2\xi}\left[k_\text{cut}-2\text{atan}(k_\text{cut}/2)\right].
		\end{split}
	\end{align}
	The ground state population then fullfills
	\begin{align}
		N_0=N_C-\frac{2\text{{$L_d$}}^3}{\pi\xi^3}\frac{\xi^2}{\lambda_\text{th}^2}\left[k_\text{cut}-\text{atan}\left(\frac{k_\text{cut}}{2}\right)\right].
	\end{align}
	
	\begin{figure}
		\centering
		\includegraphics[width=1\linewidth]{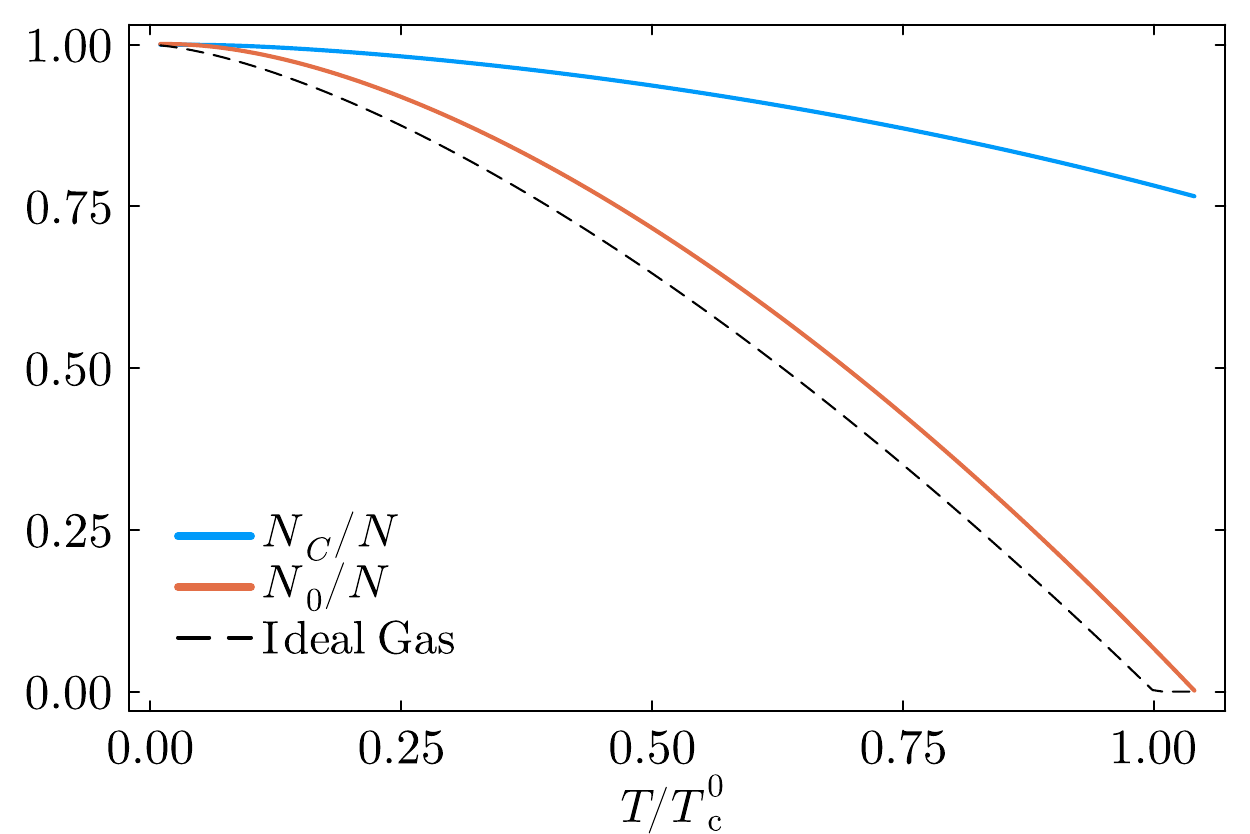}
		\caption{Occupation of the coherent region (blue solid) and of the ground state (orange solid) against the temperature $T/T_\text{c}^0$ (with $T_c^0$ the ideal gas critical temperature) for parameters $N_\text{tot}=10^5$, $\text{{$L_d$}}=27\mu$m, $a_\text{s}=25$nm, $m=6.5\cdot10^{-26}$kg {($^{39}$K)}, correpsonding to the density, mass and scattering length in \cite{hilker_first_2022}. Most of the particles populate the coherent region already for temperatures above the critical temperature. Compared to the ideal gas (dashed) the ground state occupation is enhanced by the interaction\cite{gruter_critical_1997,stoof_nucleation_1992}.}
		\label{fig:groundstateccupation}
	\end{figure}
	Figure \ref{fig:groundstateccupation} shows the coherent region and ground state occupation for represantative parameters.
	
	\section{Error Estimation}
\begin{figure}
		\centering
		\includegraphics[width=1\linewidth]{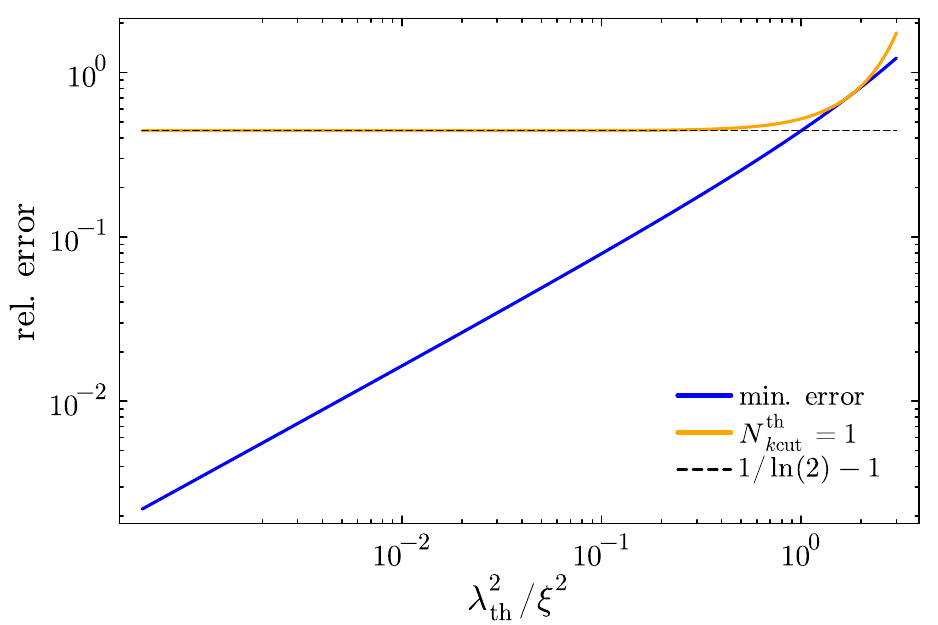}
		\caption{Estimate of the relative error of the highest occupied state in the coherent region (determined by its deviation from the Bose-Einstein statistics (\ref{thoccupation})). The orange line displays the rule of thumb $N_{k\text{cut}}^\text{th}=1$, while the blue line shows the error for the optimised cut-off choice (\ref{cut-offchoice}). While only slightly deviating for $\lambda_\text{th}^2/\xi^2\lesssim1$, at higher temperatures the optimized choice greatly reduces the error.}
		\label{fig:ncut1err}
	\end{figure}

	As the states in the coherent region should be \emph{highly} populated, a rule of thumb asking for $N_{k\text{cut}}^\text{th}=1$ is often applied\cite{comaron_quench_2019,roy_finite-temperature_2021,keepfer_phase_2022,roy_finite-temperature_2023,krause_thermal_2024}. While a nearly optimal choice for low temperatures $\lambda_\text{th}^2/\xi^2\lesssim1$, at higher temperatures the SPGPE can be expected to give more reliable results by choosing a smaller cut-off (see figure \ref{fig:ncut1err}).
	
	\section{Number damping bound}
	In section \ref{Decay} we make use of an upper bound for the strength of number damping derived in the following. It is
	\begin{align}
		\begin{split}
		\gamma&=\frac{8a_\text{s}^2}{\lambda_\text{th}^2}e^{\beta\mu_{3\text{D}}}\int_0^1dy\ln\left(\frac{1-zy}{1-z}\right)\frac{1}{(1-y)(1-zy)}\\
		&=\frac{8a_\text{s}^2}{\lambda_\text{th}^2}e^{\beta\mu_{3\text{D}}}\frac{1}{1-z}\int_0^{z/(1-z)}dx\frac{\ln(1+x)}{x(1+x)}\\
		&<\frac{8a_\text{s}^2}{\lambda_\text{th}^2}\frac{e^{\beta\mu_{3\text{D}}}}{1-z}\frac{\pi^2}{6},
		\end{split}
	\end{align}
	where we substituted $x=z(1-y)/(1-z)$. From the second to the third line we used that the integrand is positive to obtain an upper bound by setting the upper integration limit to infinity. We have further
	\begin{align}
		\begin{split}
		\frac{e^{\beta\mu_\text{3D}}}{1-z}&=\frac{e^{2\beta(\epsilon_\text{cut}-\mu_\text{3D})}}{e^{\beta(2\epsilon_\text{cut}-\mu_\text{3D})}-1}\\
		&<\frac{e^{2\beta(\epsilon_\text{cut}-\mu_\text{3D})}}{e^{2\beta(\epsilon_\text{cut}-\mu_\text{3D})}-1}\\
		&=\frac{(N_\text{cut}+1)^2}{2N_\text{cut}+1}\\
		&<\frac{N_\text{cut}}{2}+1.
		\end{split}
	\end{align}
	We conclude the bound
	\begin{align}
		\label{gambound}
		\frac{\gamma\lambda_\text{th}^2}{8a_\text{s}^2}<\frac{\pi^2}{6}\left(1+\frac{N_\text{cut}}{2}\right)
	\end{align}
	used in the main text.

\end{document}